\documentclass[reprint]{revtex4-2}

\usepackage{graphicx}
\usepackage{dcolumn}
\usepackage{bm}

\usepackage[utf8]{inputenc}
\usepackage[T1]{fontenc}
\usepackage{mathptmx}
\usepackage{etoolbox}
\usepackage{amsmath}

\usepackage[separate-uncertainty = true]{siunitx}

\makeatletter
\def\@email#1#2{%
 \endgroup
 \patchcmd{\titleblock@produce}
  {\frontmatter@RRAPformat}
  {\frontmatter@RRAPformat{\produce@RRAP{*#1\href{mailto:#2}{#2}}}\frontmatter@RRAPformat}
  {}{}
}%
\makeatother

\begin{document}

\title{Acoustic monitoring of the gelation of a colloidal suspension} 

\author{Nicolas B{\'e}licard}
\email[]{nicolas.belicard@gmail.com}

\author{Marc Junior Ni{\'e}met-Mabiala}

\author{Jean-Noel Tourvieille}
\email[]{jean-noel.tourvieille@solvay.com}

\author{Pierre Lidon}
\email[]{pierre.lidon@u-bordeaux.fr}

\affiliation{Univ. Bordeaux, CNRS, Solvay, LOF, Pessac, F-33600, France}

\date{\today}

\begin{abstract}
Because they are sensitive to mechanical properties of materials and can propagate even in opaque systems, acoustic waves provides us with a powerful tool for local rheological characterization of various systems. While most common acoustic techniques rely on time-of-flight measurements, acoustic spectra of speed and attenuation contain rich information on the propagation medium which led to the development of acoustic spectroscopy techniques. They however remain underused in the field of complex fluids because of the difficulty to interpret quantitatively acoustic signals. In this note, we use a simple ultrasound spectroscopy setup to investigate the gelation dynamics by precipitation of a silica colloidal gel. First, we show that a simple analysis of acoustic attenuation allows to define a gelation time, which is proportional to this obtained with rheological measurements. This validates the possibility to use acoustic spectroscopy for monitoring of gelation process, and our setup also has the possibility to perform mappings, showing that the formed gels display some heterogeneity. By studying in more details acoustic spectra, we finally attempt to relate more precisely acoustic measurements with mechanical parameters of the material.
\end{abstract}

\pacs{}

\maketitle 

\section{Introduction}
\label{sec1}

Acoustic waves are coupled perturbation of deformation and pressure propagating across any material medium (even opaque), that can be longitudinal (associated with compression) or transverse (associated with shear deformations, strongly damped in liquids). Acoustic methods are thus interesting characterization tools, alternative or complementary to their optical counterparts, giving access to different parameters and systems ranging from industry to biomedical research~\cite{bercoff_2004,errico_2015,wiklund_2007,gallot_2013,mograne_2019}. In particular, the short wavelengths associated with high-frequency ultrasonic waves offer a satisfactory spatial resolution, typically around a few hundreds of microns in aqueous media. Simple imaging approaches rely essentially on the measurement of time-of-flight of acoustic echoes in heterogeneous system, and thus on sound speed: however, acoustic attenuation also contains rich information. 

In simple, homogeneous liquids, heat conduction effects on sound propagation are negligible and attenuation is mostly determined by shear viscosity, bulk viscosity and possible relaxation phenomena between molecular energy levels~\cite{markham_1951,hirai_1958}. Acoustic spectroscopy thus appears as an insightful analytic technique, complimentary to more conventional optical methods by giving access to new materials, and commercial devices have been developed to measure sound speed and attenuation spectra. This gave rise to fine studies of microscopic dynamics in simple liquids for instance~\cite{eggers_1996,kaatze_2000}.

In complex fluids, relating acoustic spectra to physical parameters is more challenging. In viscoelastic materials, one can define viscoelastic acoustic moduli, analogous to their rheological counterparts despite the very distinct frequency ranges probed by ultrasound and conventional rheometers~\cite{longin_1998,verdier_1998,leroy_2010,scanlon_2015,lefebvre_2018}. In heterogeneous materials, acoustic propagation is also affected by the shape and by the size distribution of the dispersed phase. For instance, in well controlled suspensions, measured  attenuations~\cite{forrester_2016a,forrester_2016b,mori_2018} can be described by sophisticated models~\cite{epstein_1953,allegra_1972,challis_2005,allashi_2014,pinfield_2015,valierbrasier_2015,forrester_2019} or by more empirical approaches~\cite{dukhin_1996b,dukhin_characterization}. In spite of these efforts, modeling acoustic spectra remains out of reach for most complex fluids, aside from a few model systems. 

Even without a quantitative interpretation of acoustic spectra, ultrasound spectroscopy remains an interesting tool for complex material characterization and has been occasionally used since the 1950es. By first establishing correlations between acoustic measurements and parameters of interest, it is possible to characterize semi-quantitatively complex systems. Acoustic speed is for instance particularly easy to measure and allows to access elastic properties of the studied material~\cite{maleky_2007,maleky_2011}. Moreover, average attenuation, or preferably full attenuation spectra~\cite{franco_2019}, can be used as a proxy for viscosity in many systems (e.g. in food science~\cite{giordano_1983,dalgleish_2005,kuo_2008,resa_2009,laux_2013,laux_2014,derra_2018}, biomedical ~\cite{landini_1986,jongen_1986,parker_1988} or pharmaceutical~\cite{bonacucina_2008,bonacucina_2016} applications). It is then also possible to monitor physico-chemical transformations in time~\cite{dalgleish_2005,krasaekoopt_2005,resa_2009,derra_2018}. Such an approach offers promising perspectives for the study of industrial systems. In particular, the technique is easy and cheap to setup, possibly for online monitoring. It gives access to opaque systems, out of reach of common optical techniques. Finally, because of the vanishingly small deformations associated to acoustic wave propagation, it causes minimal disturbance to the system which can be of high interest for instance for thixotropic materials with long relaxation times or fragile materials with very limited linear rheological regime. Despite these promises, acoustic spectroscopy remains underused in the physico-chemistry community.

As they are associated with changes of microstructure and mechanical parameters, gelation processes appear as good candidates for acoustic monitoring. Such studies have for instance been performed with various polymer gels~\cite{bacri_1980,sidebottom_1993,parker_2012}. Colloidal gels constitute a more complex case, as they are both viscoelastic and dispersed materials, and acoustic studies focused on model silica gels obtained by sol-gel processes. They showed a qualitative correlation between rheological measurements and the observed increase of sound speed and attenuation~\cite{serfaty_1998,forest_1998,cros_2001,griesmar_2003,ouldehssein_2006,robin_2012}. All these systems belong to the group of chemical gels, where mesoscopic elements are linked through chemical bonds.

Surprisingly, despite their omnipresence in nature and industry and thorough rheological study~\cite{moller_2006,coussot_2014,bonn_2017}, no acoustic study of physical gelation processes has been undertaken. In this work, we propose to bridge this gap. We use a simple setup of acoustic spectroscopy which measures sound speed and attenuation spectra during the gelation process. We decided to study the gelation of a colloidal silica suspension induced by aggregation caused by the screening of surface charges by a concentrated brine, leading to a weak gel which can be heterogeneous~\cite{kurokawa_2015}. This system has already been studied in details through optical and rheological techniques~\cite{trompette_2003,trompette_2004,cao_2010} that we use as a guideline for formulation and a reference to compare with our measurements. The formulation of the gel, rheological characterization and the description of the acoustic setup are gathered in Section~\ref{sec2}. 

In Section~\ref{sec3}, we first present rheological characterization of the gelation process. From monitoring of linear viscoelastic moduli over time, we define several gelation times which are of similar order of magnitude and show the same exponential evolution with brine concentration as reported in previous studies. We then propose a simple analysis of acoustic signals: while sound speed do not significantly evolve during gelation, acoustic attenuation continuously increases.Empirically fitting the average attenuation allows us to define an acoustic gelation time, which correlates well with the rheological observations. This proves that our simple acoustic method is representative of the dynamics of mechanical evolution of the gel. We also exploit the possibility to scan the measurement cell with the acoustic beam to map the acoustic properties of a gel in stationary state: we observe some heterogeneities, which could partially explain the sample to sample variability of our measurements. 

Finally, in Section~\ref{sec4}, we attempt to relate more precisely our acoustic measurements with rheological parameters by analyzing more carefully the attenuation spectra and their evolution in time, by using two different approaches. First, we compute acoustic viscoelastic moduli and show that viscous modulus allows to define a frequency-dependent characteristic time by an empirical exponential fit. Then, we propose a more physical model considering the attenuation results from the surperposition of this due to the solvent and this due to a single time scale relaxation phenomenon, modeling the effect of the elastic backbone. We finally discuss the compatibility of these two approaches.



\section{Materials and methods}
\label{sec2}

\subsection{Formulation of Ludox gels}
\label{subsec2_1}

In this work, we focus on the same type of gel as in the studies by Trompette et al.~\cite{trompette_2000,trompette_2003,trompette_2004,trompette_2017} and by Cao et al.~\cite{cao_2010} that served as a guide for formulation. In all experiments, we used an initial suspension of silica nanoparticles (Ludox HS-40, Sigma Aldrich) with a $\SI{40}{\percent}$wt. content of solid and average particle radius $\SI{6}{\nano\meter}$ as specified by the manufacturer. The solution is mixed with a brine made of sodium chloride ($\mathrm{NaCl}$) salt dissolved in deionized water. In order to tune gelation kinetics, we modified the brine concentration between $\SI{0.35}{\mole\per\liter}$ and $\SI{0.65}{\mole\per\liter}$ while keeping a similar volume fraction $\varphi=\SI{5.9}{\percent}$ of Ludox particles (assuming a density of $\rho=\SI{1.3}{\kilo\gram\per\meter\cubed}$ for the initial suspension, as indicated by the supplier) in the final gel.

Physico-chemical properties of Ludox silica colloidal suspensions are complex~\cite{hallez_2017,trompette_2017} but can be qualitatively described by DLVO theory. In the initial suspension, the negative surface charge of particles induces a dominant electrostatic repulsion which stabilizes the solution. When mixed with brine, salts in solution screen this surface charge and attractive van der Waals interaction dominates the interaction potential, leading to a progressive agglomeration of colloids. A fractal network of particles forms and grows in time, giving elastic properties to the gel when it percolates across the sample. It is important to note that the obtained gels are physical and thus a priori distinct in microstructure and properties from the chemical gels obtained with the sol-gel process, which have already been studied by acoustic spectroscopy.

Both for acoustical and rheological measurements, the system is maintained at a constant temperature $T=\SI{23.0(1)}{\celsius}$ and the brine and Ludox suspensions are initially thermalized at the same temperature, in order to limit temperature transients. The two solutions are vigorously introduced in the measuring cell to favor an initial homogeneous mixing which sets the origin of time $t = 0$ in the experiments. The solution progressively turns into a gel, which can be visually observed by an increasing turbidity due to the growth of the colloidal network. The volume fraction of Ludox and concentration of the brine were chosen to keep a gelation time long enough to allow for loading and thermalization of the sample, while being short enough to ensure experiments run over less than one day.

\subsection{Rheological measurements}
\label{subsec2_2}

In order to have a basis for interpretation of acoustic signals, we performed rheological characterizations by measuring the evolution of the viscoelastic moduli during the gelation process. Experiments are done in a sand-blasted cylindrical Couette geometry (inner radius~$\SI{25}{\milli\meter}$, gap~$\SI{2.5}{\milli\meter}$) with a stress-controlled rheometer (Malvern Kinexus Extra+), at controlled temperature $T=\SI{23}{\celsius}$.

After loading the suspension and the brine, we monitor the evolution over time of the viscoelastic moduli at an imposed shear amplitude of $\gamma=\SI{0.1}{\percent}$, in order to minimally disturb the gelation process and remain in the linear rheological domain after gelation for all probed frequencies. Gelation is slow enough to allow for measurements at different frequencies: rheometer repeats over time seven steps of frequencies ranging from $\SI{0.1}{\hertz}$ to $\SI{1}{\hertz}$.

\subsection{Acoustic setup and measurements}
\label{subsec2_3}

\subsubsection{Setup}
\label{subsec2_3_1}

\begin{figure*}[!htb]
\centering
\includegraphics{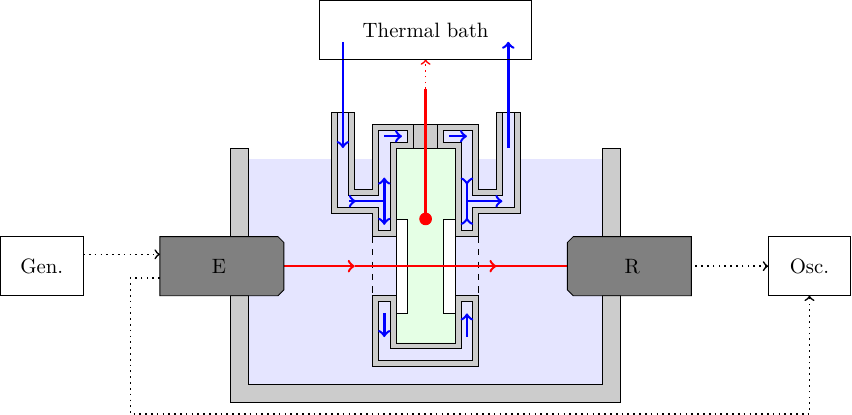}
\caption{Schematics of the acoustic setup. The sample under study (in green) is placed in a double walled cell, allowing inner water circulation (blue arrows) to control the temperature through a feedback loop driven by a thermometer plunged in the sample. The cell is closed by polycarbonate windows (in white) to allow ultrasound propagation. Acoustic transducers (dark gray) are placed and aligned on both sides of the cell. The emitting transducer E is powered by a pulse generator (Gen.) and is also used to acquire reflected signals. The receiving transducer T is used to collect transmitted signals. Both signals are digitized by an oscilloscope (Osc.). The acoustic path is materialized by the red line.}\label{fig:setup}
\end{figure*}

The setup is depicted on Fig.~\ref{fig:setup}. Two piezoelectric immersion transducers (Olympus V312-SU, central frequency~$\SI{10}{\mega\hertz}$, bandwidth~$4-\SI{12}{\mega\hertz}$, diameter~$\SI{6}{\milli\meter}$) were placed on both sides of the sample, in a small water tank in order to maximize acoustic transmission. One transducer is connected to a pulse generator (Sofranel DPR300) which generates short electrical impulses (duration around~$\SI{0.5}{\micro\second}$) with a tunable repetition rate in order to create acoustic pulses. Our setup allows for two different measurements: the emitting transducer can also be used as a receiver to measure reflected signals while the other transducer is used to acquire transmitted signals. The earlier was connected to an amplification and filtering stage included in the pulse generator, and both are eventually linked to an oscilloscope (Keysight InfiniiVision 2000X) triggered by the pulse generator for data acquisition and digitization. In order to improve the signal-noise ratio, measurements were averaged over ten pulses with a pulse repetition frequency of $\SI{1}{\hertz}$ to ensure a good separation between successive pulses while being significantly faster than the typical timescale of evolution of the gel.

We took great care to ensure the sample was kept under isothermal conditions to avoid any thermal bias. To this purpose, the sample was placed in a double-walled cell built in a resin by a SLA 3D printer (Formlabs, High Temp resin). The double wall allowed circulation of water whose temperature is imposed by a thermal bath driven by a feedback loop controlled by a platinum resistance thermometer (4 point PT100, Radiospares) plunged inside the sample. We managed to obtain temperature fluctuations below $\SI{0.1}{\celsius}$ after typical thermalization transients of a few minutes.

In order to limit interfaces on the acoustic path, the cell wall was designed with a small opening on both sides which is closed by a polycarbonate window (in white on Fig.~\ref{fig:setup}) of homogeneous thickness about $\SI{1}{\milli\meter}$, which was isotropic and homogeneous for ultrasound propagation.

In order to align the setup, the measurement cell was held by a ball joint on translation platforms in order to control both its orientation and position. First, the two transducers were aligned parallel and along the same axis by maximizing the amplitude of the transmitted signal. Then, the cell was placed in between and aligned with windows orthogonal to the acoustic propagation direction by maximizing the amplitude of the signal reflected by the first window.

\subsubsection{Signal analysis}
\label{subsec2_3_2}

We assume that acoustic signals can be described by unidimensional wave packets in our setup. More precisely, the beams generated by the transducers are axisymmetric and we assume that the transverse structure of the acoustic field is not significantly affected by the gelation inside the experimental cell. As we ultimately normalize measurement by a reference pulse propagating in water, effects of this transverse structure factor out in our final results~\cite{rogers_1974}. 

The acoustic pressure at position $x$ and time $t$ for a single freely propagating pulse can then be written as a wave packet:
\begin{equation}
u(x,t) = \int u^0_f e^{j \left[k(f)x - 2\pi f t \right]} \mathrm{d} f \equiv \int u_f (x) e^{-2 j \pi f t} \mathrm{d}f
\end{equation}
\noindent where $u^0_f$ represents the spectral content of the source and $u_f$ is the Fourier transform of the signal. The wave vector $k(f) = k'(f) + j k''(f)$ is a complex number, which is set for a given frequency by the properties of the material through a dispersion relation. One defines the (frequency-dependent) sound speed $c_f$ and attenuation coefficient $\alpha_f$ by setting $k' = f/c_f$ and $k''=\alpha_f$. For the sake of simplicity, we will from now on drop the subscript $f$ for speed and attenuation, that will be simply denoted by $c$ and $\alpha$: yet, one should keep in mind that these quantities depend on frequency. The Fourier transform of the wave packet can then be expressed as
\begin{equation}
u_f (x) = u^0_f e^{2 j \pi f x / c} e^{-\alpha x}.
\end{equation}
\noindent For propagation over a distance $\ell$ of a wave at frequency $f$, the time of flight is given by $\tau = \ell/c$ and the attenuation factor is $e^{-\alpha \ell}$. With this definition, the attenuation $\alpha$ is expressed in Neper per meter ($\SI{}{Np\per\meter}$).

When a wave packet crosses an interface, it generates a transmitted and a reflected echo of proportional amplitudes, that later propagate towards the next interface. In our setup, the two transducers allowed us to record these successive echoes created by the initial pulse during its propagation in the water tank, in the windows of the cell and in the sample of interest. The time of flight and attenuation of these pulses were thus not only related to the sample, but also to the propagation in the whole setup: obtaining absolute measurements is not possible with our current experiment. We can however obtain relative values with respect to a reference medium, chosen as water because its acoustic properties are precisely known.

The calibration procedure is described in details in~\ref{A_sec1}. Shortly, we normalized our measurements by reference signals obtained by reflection on a cell filled with air (the air/window interface being perfectly reflecting for acoustic waves) and by reflection on and transmission through a cell filled with water. This allowed us to obtain the reflection and transmission coefficients of the various interfaces and to factor out the effects of propagation through other media. We finally obtained the excess time of flight and excess attenuation for the propagation across the length $\ell$ of sample, compared to the propagation across a similar length of water. As the length $\ell$ was also calibrated, and by using tabulated acoustic properties of water~\cite{marczak_1997,holmes_2011}, we could deduce the sound speed and attenuation of the sample.



\section{Correlation between rheological and acoustical results}
\label{sec3}

\subsection{Gelation time(s) from rheological measurements}
\label{subsec3_1}

\begin{figure*}[!htb]
\centering
\includegraphics{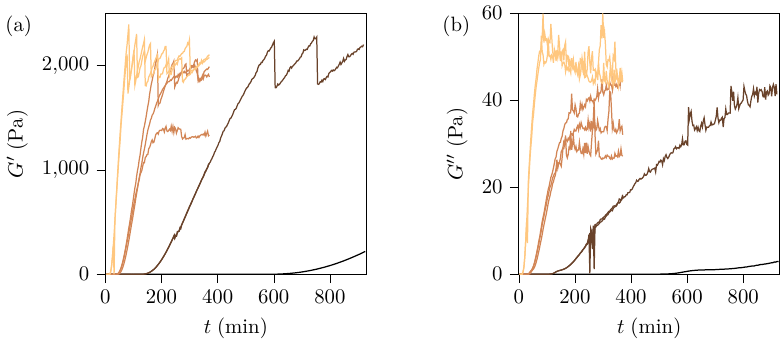}
\caption{Evolution of the elastic (a) and viscous (b) moduli at frequency $f=\SI{1}{\hertz}$ and shear amplitude $\gamma=\SI{0.1}{\percent}$ during the gelation. For all samples, the volume fraction of particles is $\varphi=\SI{5.9}{\percent}$. Lighter colors correspond to more concentrated brines, and brine concentrations are respectively $[\mathrm{NaCl}] = 0.35$, $0.45$, $0.55$ and $\SI{0.65}{\mole\per\liter}$.}\label{fig:rheol_Gp_Gs_vs_t}
\end{figure*}

The evolution in time of the viscoelastic moduli $G'$ and $G''$ after mixing the brine and the colloidal suspension are presented in Fig.~\ref{fig:rheol_Gp_Gs_vs_t}, for different brine concentrations and similar particle volume fraction. They are reported here at a single frequency of $\SI{1}{\hertz}$ but our measurements showed no notable difference between the different probed frequencies. The overall evolution is consistent with results reported by Cao et al.~\cite{cao_2010}. Initially, both moduli are very small and at the limit of resolution of the rheometer. After a time delay $\tau_\text{d}$, they suddenly rise, and eventually increase more progressively as the gel ages. For some samples, we observed sudden drops of the elastic modulus when the system was gelified: we suspect that this is due to wall slip events, despite the walls of the geometry were sandblasted. As they happen after gelation, these events do not affect our measurements of gelation time. Moreover, as it is more directly related to acoustic attenuation, we will mostly focus on measurements obtained from the viscous modulus $G''$ that does not display these drops.

In order to check for reproducibility, we repeated some of the formulations: despite some variability in the absolute values of the moduli (particularly visible on their final values), there is still a clear difference of dynamics when varying the amount of salt in the system, more concentrated brine leading to faster gelation. However, it is to note that the least concentrated sample (for $[\mathrm{NaCl}]=\SI{0.35}{\mole\per\liter}$) hardly started to gelify at the end of the experiment: it is thus dubious to interpret results for this gel. We will present our measurement for this gel on figures, but it will be excluded for fittings. Similar caution will be taken with acoustic measurements.

Different ways exist to define a gelation time from rheological measurements. For all the different definitions presented in the following, the relative variation of gelation time obtained when reproducing the experiment in the same conditions is about $\SI{10}{\percent}$, which is taken as the associated uncertainty.

The most rigorous definition is given by the Winter criterion: by analyzing the evolution in time of the viscoelastic moduli at different frequencies, the gelation time $\tau_\text{W}$ can be defined as the time at which the loss angle $\tan\delta = G'' / G'$ is independent of frequency. This criterion was justified by theoretical arguments on the behavior of viscous and elastic moduli when approaching the critical point of percolation~\cite{winter_1987,chambon_1987,holly_1988,hodgson_1990,winter_1991}. While being initially introduced for chemical gels forming by reticulation, this criterion has also been successfully applied to physical silica gels~\cite{trompette_2000,ponton_2002}. The time defined from Winter criterion also coincides approximately with the time at which the elastic modulus $G'$ overcomes the viscous one $G''$~\cite{drabarek_2002,matsunaga_2007}, which is a more intuitive and common definition. It is interesting to note that applying this latter definition would allow to define a frequency-dependent gelation time but our rheological measurements showed almost no effect of frequency on the probed range.

\begin{figure*}[!htb]
\centering
\includegraphics{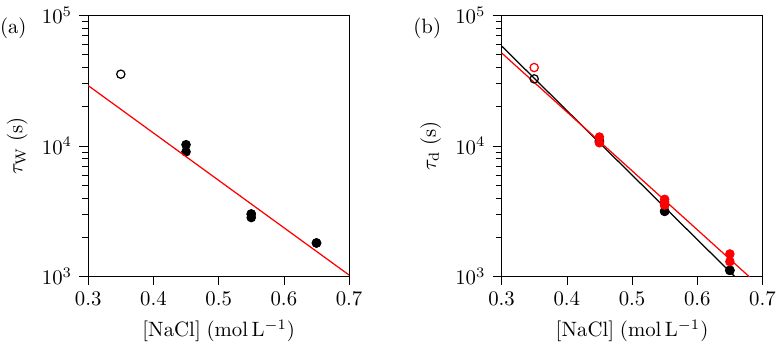}
\caption{(a) Gelation time obtained from Winter criterion for varying brine concentrations. Red line displays an exponential fit along Eq.~(\ref{eq:tau_exp_conc}) with $\tau_{0,\text{W}} \simeq \SI{3.6e5}{\second}$ and $\kappa_\text{W} = \SI{8.5(5)}{\liter\per\mole}$. (b) Gelation time obtained from the sudden rise of elastic (black symbols) and viscous (red symbols) moduli for varying brine concentrations. Lines correspond to an exponential fit along Eq.~(\ref{eq:tau_exp_conc}) with $\tau_{0,\text{d}} \simeq \SI{1.8e6}{\second}$ and $\kappa_\text{d} = \SI{11.4(5)}{\liter\per\mole}$ (black line), and $\tau_{0,\text{d}} \simeq \SI{1.1e6}{\second}$ and $\kappa_\text{d} = \SI{10.4(5)}{\liter\per\mole}$ (red line). In all fits, the point at $[\mathrm{NaCl}]=\SI{0.35}{\mole\per\liter}$ (open symbol) has been excluded.}\label{fig:rheol_gelation_time}
\end{figure*}

Measurements by Cao et al.~\cite{cao_2010} showed that gelation time obtained from Winter criterion evolves exponentially with concentration $[\mathrm{NaCl}]$ along: 
\begin{equation}
\tau = \tau_{0} e^{-\kappa[\mathrm{NaCl}]}
\label{eq:tau_exp_conc}
\end{equation}
\noindent with $\kappa_\text{litt} = (8.3 \pm 1.2) \SI{}{\liter\per\mole}$ for gels similar to the one we study (in particular with similar volume fraction of particles). The times obtained from Winter criterion for different brine concentrations are plotted on Fig.~\ref{fig:rheol_gelation_time}(a). They follow this exponential evolution, with $\tau_{0,\text{W}} \simeq \SI{3.6e5}{\second}$ and $\kappa_\text{W} = \SI{8.5(5)}{\liter\per\mole}$. We observe that the coefficient $\kappa$ we obtain is in quantitative agreement with the value obtained in the literature.

Winter criterion is not very convenient to use. First, it was difficult with some samples to observe a clear crossing of the $\tan\delta$ curves at different frequencies. Moreover, these measurements are long to perform as they require to be made at different frequencies. It is thus interesting to define gelation times by other means. 

A simple criterion is obtained by measuring the delay time $\tau_\text{d}$ at which the viscoelastic moduli start rising: the obtained results are displayed on Fig.~\ref{fig:rheol_gelation_time}(b). We observe that the times obtained by analyzing the elastic and viscous moduli are similar: consequently, we will mostly focus on the times obtained from $G''$, as they are more likely to be correlated with the acoustic attenuation that will be exploited in section~\ref{subsec3_2}. This time also follows an exponential evolution with concentration~(\ref{eq:tau_exp_conc}), with a typical time $\tau_{0,\text{d}} \simeq \SI{1.1e6}{\second}$ and concentration $\kappa_\text{d} = \SI{10.4(5)}{\liter\per\mole}$. This value of $\kappa$ is close to this obtained from Winter criterion, suggesting that the delay time is also a good marker of gelation.

We also tested other ways to define a gelation time, by measuring a rising time at $\SI{63}{\percent}$ of the final value and by fitting viscoelastic moduli with a stretched exponential, as suggested by Cao et al.~\cite{cao_2010} for the elastic modulus. In all cases, we obtain comparable times from studies of $G'$ and $G''$, independent on frequency, and displaying an exponential variation with brine concentration along Eq.~\eqref{eq:tau_exp_conc} with similar values of $\kappa$. 

In the following, we will rather discuss the time observed from Winter criterion. However, the fact that the other proposed definition give qualitatively similar results makes us confident in the quality of our measurements of gelation time.

\subsection{Gelation time from averaged attenuation}
\label{subsec3_2}

\begin{figure}[!htb]
\centering
\includegraphics{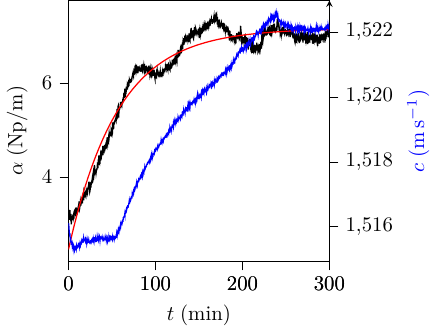}
\caption{Example of the acoustic attenuation $\alpha$ (black curve, left axis) and sound speed (blue curve, right axis) evolution during the gelation of a gel, with initial brine concentration $[\mathrm{NaCl}] = \SI{0.55}{\mole\per\liter}$. The red curve corresponds to an exponential fit of the attenuation $\alpha$ according to Eq.~(\ref{eq:attenuation_exp_model}) with $\Delta \alpha = \SI{4.7}{Np\per\meter}$, $\alpha_0 = \SI{2.4}{Np\per\meter}$ and $\tau_\text{US}=\SI{3.4e3}{\second}$. }\label{fig:US_example}
\end{figure}

As a first step, we want to compare the gelation dynamics probed by ultrasound spectroscopy to this obtained in rheology. To do so, we decided to focus on a simple analysis of acoustic data and consider only the averaged sound speed $c$ and attenuation $\alpha$ on the whole frequency range. Information is obviously lost in the process and we will check in Section~\ref{sec4} that it still makes sense, but this approach is interesting from an engineering point of view, as it provides us with a simple method to monitor gelation. We thus reproduced the same formulations studied with the rheometer and monitored their gelation with the acoustic setup presented in Section~\ref{subsec2_3}. An example of evolution of average sound speed $c$ and attenuation $\alpha$ is displayed on Fig.~\ref{fig:US_example}. In some samples, initial few minutes were removed from analysis because of transient thermalization of the sample.

We observe that sound speed does not vary significantly over the duration of the experiment. This is probably due to the fact that it is dominated by the solvent compressibility, which does not evolve during gelation. The tiny remaining evolution (overall increase observed in this example) is not reproducible from one experiment to another and probably reflects some very local changes in the gel. We consequently will not exploit data from sound speed in the following. On the contrary, the attenuation displays a reproducible and progressive increase during the gelation process.

\begin{figure}[!htb]
\centering
\includegraphics{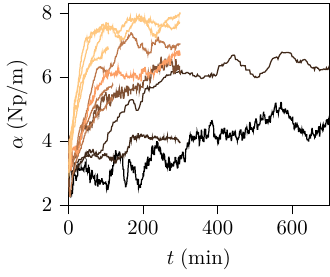}
\caption{Evolution of the acoustic attenuation $\alpha$ in time, for gels with identical particle volume fraction $\varphi = \SI{5.9}{\percent}$ and varying salt concentration $[\mathrm{NaCl}]=0.35$, $0.4$, $0.45$, $0.50$, $0.55$ and $\SI{0.60}{\mole\per\liter}$. Lighter colors correspond to more concentrated brines.}\label{fig:US_attenuation_vs_time}
\end{figure}

This evolution for the samples of different brine concentrations is plotted on Fig.~\ref{fig:US_attenuation_vs_time}. Despite some variability, attenuation overall increases in time and dynamics is faster for higher brine concentrations: this qualitative observation is consistent with the aforementioned literature on Ludox gels~\cite{trompette_2003,trompette_2004,cao_2010} and our own rheological measurements of~\ref{subsec3_1}. This is also consistent with the fact that acoustic attenuation is somehow related to dissipation in the material, thus to the viscous modulus $G''$. As in our rheological experiments, the sample at brine concentration $[\mathrm{NaCl}]=\SI{0.35}{\mole\per\liter}$ did not fully gelify over the duration of the experiment: we will display the associated result but exclude it from further analysis.

\begin{figure}[!htb]
\centering
\includegraphics{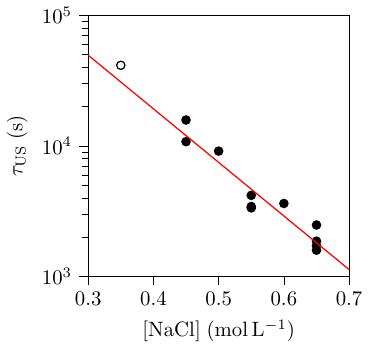}
\caption{Gelation time obtained from exponential fit of acoustic attenuation over time, for varying salt concentrations. The red line displays an exponential fit along Eq.~(\ref{eq:tau_exp_conc}) with $\tau_{0,\text{US}} \simeq \SI{8.4e5}{\second}$ and $\kappa_\text{US} = \SI{9.5(8)}{\liter\per\mole}$. The open symbol corresponds to the sample at $[\mathrm{NaCl}]=\SI{0.35}{\mole\per\liter}$ that did not fully gelify over the experiment duration and is excluded from the fit.}\label{fig:US_time_vs_conc}
\end{figure}

In order to compare our results with rheometrical characterization, we need to define a typical gelation time $\tau_\text{US}$ from the acoustic measurements. Here, there is no apparent delay in the evolution of $\alpha$ and noise in our data makes difficult to obtain a robust rising time. We propose to fit the evolution by a generic exponential function:
\begin{equation}
\alpha (t) = \alpha_0 + \Delta \alpha \left( 1 - \mathrm{e}^{-t/\tau_\text{US}} \right).
\label{eq:attenuation_exp_model}
\end{equation}
\noindent Such an evolution describes satisfactorily the evolution of attenuation, as shown in Fig.~\ref{fig:US_example}, and the resulting gelation times are plotted for varying brine concentration in Fig.~\ref{fig:US_time_vs_conc}. When reproducing the experiment with samples of similar composition, typical relative variations of $\tau_\text{US}$ are of about $\SI{20}{\percent}$.

The evolution of gelation time again follows an exponential evolution with brine concentration, given by Eq.~(\ref{eq:tau_exp_conc}) with $\tau_{0,\text{US}} \simeq \SI{8.4e5}{\second}$ and $\kappa_\text{US} = \SI{9.5(8)}{\liter\per\mole}$. 

\subsection{Perspectives for monitoring of gelation processes}
\label{subsec3_3}

As presented in section~\ref{subsec3_1} and~\ref{subsec3_2}, the gelation times $\tau_\text{W}$ or $\tau_\text{d}$  obtained from conventional rheological measurements and $\tau_\text{US}$ from acoustic measurement evolve exponentially with salt concentration with a similar characteristic concentration $1/\kappa$. This shows that these times are all proportional which suggests they all characterize gelation kinetics. This is not surprising that their values are not the same. First, the length scales that are probed by acoustics and rheometry are very different: while the rheological characterization is sensitive to the apparition of elasticity at the length scale of the rheometer gap, acoustic measurement is more local, at the scale of the acoustic beam and with a typical solicitation length scale given by the wavelength. It is interesting to note that gelation time obtained from Winter criterion, which relies on a microscopic description of the gel structure, is of the same order of magnitude as the acoustic gelation time. Moreover, the mechanical solicitation of the rheometer is pure shear while the acoustic excitation also involves compression. Finally, even if shear amplitude in the rheometer is kept as small as possible, it is likely to slightly delay the gelation process~\cite{zaccone_2009,zaccone_2010,zaccone_2011} while ultrasound amplitude is too small to have a significant impact~\cite{gibaud_2020}.

This good correlation of gelation times opens the way to a purely acoustic monitoring of gelation, of interest in industrial context as it is a non-invasive method that can easily be adapted to inline processes. In particular, there is no need for modeling of the acoustic attenuation as we used a very generic exponential fitting. It is also interesting to note that using focused transducers would allow to characterize smaller volumes of material.

In our setup, we took great care to ensure strict isothermal conditions, which would be a major shortcoming for practical applications. However, this was required in order to minimize thermal biases for validation of the method. While sound speed is known to depend strongly on temperature, acoustic attenuation is only mildly affected by temperature fluctuations and a qualitative monitoring of gelation does not require the strict temperature control we employed in this experiment.

\subsection{Mapping of acoustic properties}
\label{subsec3_4}

Our setup gives us the possibility to move the measurement cell in the acoustic field: this allows us to map the acoustic properties of of the gel. As discussed in the introduction, the gelation process of destabilization of a suspension through electrostatic screening is known to create more heterogeneous structures than the sol-gel process that has already been studied in the literature. It is thus important to check the degree of heterogeneity in the gels we study to validate our conclusions.

However, the cell presented in section~\ref{subsec2_3_1} had only a small window (in order to ensure temperature control) which did not allow for scanning a representative field of the sample. We thus needed to change the measurement cell. We formulated gels in a cell culture flask (which has the interest to offer large, flat and homogeneous sides), sealed with Parafilm to limit drying, and left them to age for about~$\SI{10}{\hour}$ in order to reach a stationary state. With this system, we could not control the temperature of the sample anymore: the flask was in a water bath to ensure acoustic propagation, but it was affected by room temperature fluctuations. Moreover, when moving the cell vertically, it was partially withdrawn from the water bath which could induce temperature changes.

\begin{figure*}[!htb]
\centering
\includegraphics{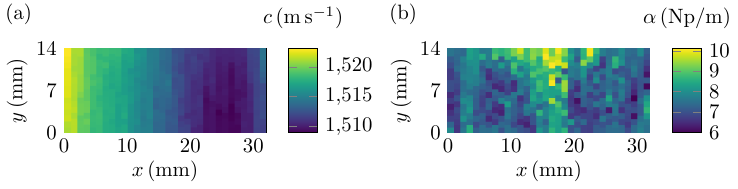}
\caption{Map of average sound speed (a) and attenuation (b) of a gel with brine concentration $[\mathrm{NaCl}]=\SI{0.55}{\mole\per\liter}$. The sample was left to age for about~$\SI{10}{\hour}$ before measurement was performed, so that it can be considered as almost stationary.}\label{fig:US_cartography}
\end{figure*}

We scanned a field of $\SI{1.4}{\centi\meter}$ in height and $\SI{3.2}{\centi\meter}$ in width, by steps of $\SI{1}{\milli\meter}$ corresponding to a map of about $450$ points. These steps are small compared to the size of the acoustic beam (of radius of about $\SI{3}{\milli\meter}$): there is thus a correlation between successive points. Such a measurement takes about $\SI{2}{\hour}$, hence the requirement to wait for a long enough time for the gel to be in a stationary state at this timescale. The cell was moved in successive vertical columns in order to limit the possible apparition of a vertical temperature gradient due to the partial withdrawal of the cell from the bath. Resulting maps of average acoustic speed and attenuation for a sample are displayed on Fig.~\ref{fig:US_cartography}. 

First, we observe an apparent lateral gradient of acoustic velocity. This is most probably due to room temperature variations which are likely on the time scale of the experiment, as there is no reason to observe such a smooth gradient orthogonal to gravity. Temperature has however a smaller impact on acoustic attenuation, as can be seen by the absence of overall tendency in acoustic maps: we can consequently still exploit these, at least qualitatively.

The result presented in Fig.~\ref{fig:US_cartography}(b) is representative of the different samples. The gel display some spots with higher attenuation in a rather homogeneous background. Without further study, it is uneasy to determine the cause of these heterogeneities: they could for instance be due to intrinsical heterogeneity of the gelation process, to imperfect initial mixing of the solutions or to the possible trapping of air bubbles. They could explain the variability of absolute values of attenuation from sample to sample. However, this does not compromise our comparative results of samples at different concentrations. These heterogeneities are indeed localized and outside of these zones, attenuation is rather homogeneous: the relative standard deviation of attenuation is there $(\sigma_\alpha / \langle \alpha \rangle) \simeq \SI{10}{\percent}$. Consequently, provided that the acoustic path crosses an homogeneous part of the sample, these fluctuations remain small enough to discriminate between the different concentrations on Fig.~\ref{fig:US_attenuation_vs_time}.



\section{Analysis of acoustic spectra}
\label{sec4}

\begin{figure*}[!htb]
\centering
\includegraphics{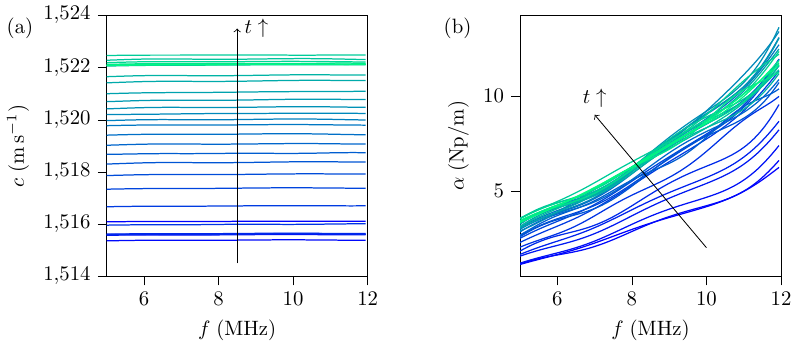}
\caption{Evolution of sound speed (a) and attenuation (b) spectra over time (lighter colors correspond to longer times, curves are spaced regularly by $\SI{4}{\minute}$) for a gel of particle volume fraction $\varphi=\SI{5.9}{\percent}$ and salt concentration $[\mathrm{NaCl}]=\SI{0.55}{\mole\per\liter}$.}\label{fig:US_spectra_time}
\end{figure*}

\begin{figure*}[!htb]
\centering
\includegraphics{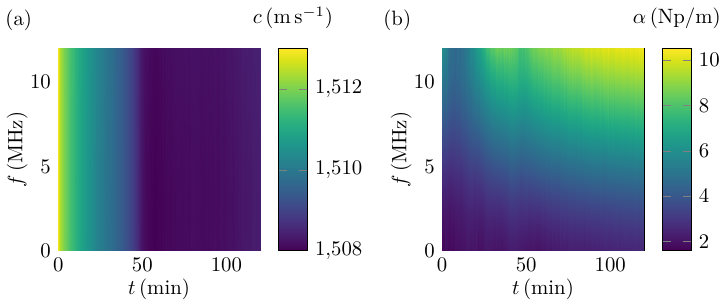}
\caption{Time-frequency representation of the acoustic speed (a) and attenuation (b) spectra during the gelation of a sample of volume fraction $\varphi=\SI{5.9}{\percent}$ and salt concentration $[\mathrm{NaCl}]=\SI{0.55}{\mole\per\liter}$. Note that the variations of sound speed are very small.}\label{fig:US_spectra_map}
\end{figure*}

The averaged approach proposed in Section~\ref{sec3} is interesting from an applied perspective as it gives a way to monitor gelation acoustically with simple data analysis. However, such an averaging over the bandwidth of the pulses loses information on the system and detailed analysis of acoustic spectra should provide further information. In this paragraph, we study these spectra and attempt to model them, in order to relate acoustic measurements with more physical parameters of the material. An example of the evolution of the acoustic spectra is presented in Fig.~\ref{fig:US_spectra_time}. An alternative representation can be obtained with colormaps in the time-frequency plane, as presented in Fig.~\ref{fig:US_spectra_map}.

We observe that there is no significant acoustic dispersion in the gel as the sound speed spectrum remains essentially flat over the whole transformation, in agreement with the idea that it is principally determined by water. However, the acoustic attenuation increases with frequency. As this evolution is monotonous, the average value remains a meaningful parameter to describe attenuation, which justifies the averaged approach, but more information could be obtained by further analysis.

\subsection{Evolution of acoustic viscoelastic moduli}
\label{subsec4_1}

\begin{figure*}[!htb]
\centering
\includegraphics{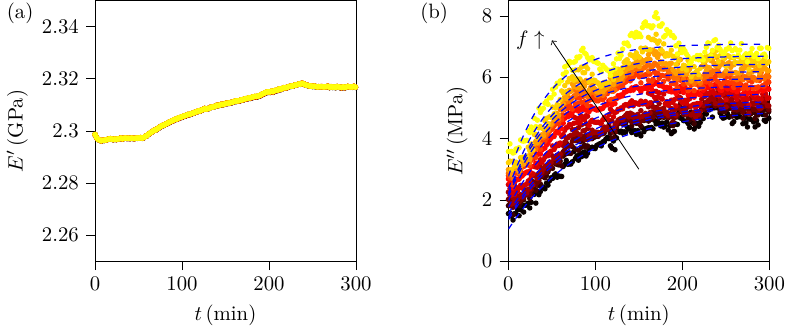}
\caption{Example of evolution of viscoelastic acoustic moduli with time, at different frequencies (from $\SI{5}{\mega\hertz}$ to $\SI{12}{\mega\hertz}$ regularly spaced by $\SI{0.8}{\mega\hertz}$, increasing from dark to bright colors) for a sample of volume fraction $\varphi=\SI{4.9}{\percent}$ and salt concentration $[\mathrm{NaCl}]=\SI{0.55}{\mole\per\liter}$. (a) Elastic acoustic modulus obtained from Eq.~\eqref{eq:definition_modulus_Epr}. (b) Viscous acoustic modulus obtained from Eq.~\eqref{eq:definition_modulus_Esec}. Dashed blue lines represent exponential fits using Eq.~\eqref{eq:exponential_fit_modulus_Esec}.}\label{fig:US_spectra_modulus}
\end{figure*}

Sound speed and attenuation are not directly related to rheological parameters. In order to rewrite these acoustic quantities in more mechanical terms, it is possible to define an elastic acoustic modulus along:
\begin{equation}
E' = \rho f^2 c^2 \frac{f^2 - (\alpha c)^2}{\left[f^2 + (\alpha c)^2\right]^2} \underset{\alpha c \ll f}{\simeq} \rho c^2,
\label{eq:definition_modulus_Epr}
\end{equation}
\noindent and a viscous elastic modulus along:
\begin{equation}
E'' = 2 \rho f^3 c^2 \frac{\alpha c}{\left[f^2 + (\alpha c)^2\right]^2} \underset{\alpha c \ll f}{\simeq} 2 \rho c^2  \cdot \frac{\alpha c}{f}.
\label{eq:definition_modulus_Esec}
\end{equation}
\noindent Moduli $E'$ and $E''$ are the analogous for acoustic propagation of the viscoelastic moduli $G'$ and $G''$ in rheology, but they are associated with different types and frequencies of excitation~\cite{dukhin_2010}. In our experiments, we have $\alpha c \ll f$ over the whole frequency range so the first order expansions in Eqs.~\eqref{eq:definition_modulus_Epr} and~\eqref{eq:definition_modulus_Esec} are valid. An example of the evolution of these moduli with time at different frequencies are displayed on Fig.~\ref{fig:US_spectra_modulus}.

As can be seen from the approximation in Eq.~\eqref{eq:definition_modulus_Epr},  the elastic modulus $E'$ is essentially related to sound speed, thus has only little dependency in frequency and does not vary significantly in time, the observed slight variation being not reproducible. The obtained value of $E' \simeq \SI{2}{\giga\pascal}$ corresponds roughly to this of water which is not surprising because elastic shear modulus of the gel is negligible compared to water bulk compressibility. On the contrary, the viscous modulus $E''$, which is mostly determined by sound attenuation both increases with time and frequency.

Following a similar empirical approach to this used with the average attenuation in~\ref{subsec3_2}, we can fit the time evolution of the viscous modulus $E''$ at every frequency with an exponential:
\begin{equation}
E'' (t)= E_0 + \Delta E \left(1-\mathrm{e}^{-t/\tau} \right).
\label{eq:exponential_fit_modulus_Esec}
\end{equation}
\noindent We thus define for every frequency a characteristic time $\tau$, as well as initial elastic modulus $E_0$ and variation amplitude $\Delta E$.

\begin{figure*}[!htb]
\centering
\includegraphics{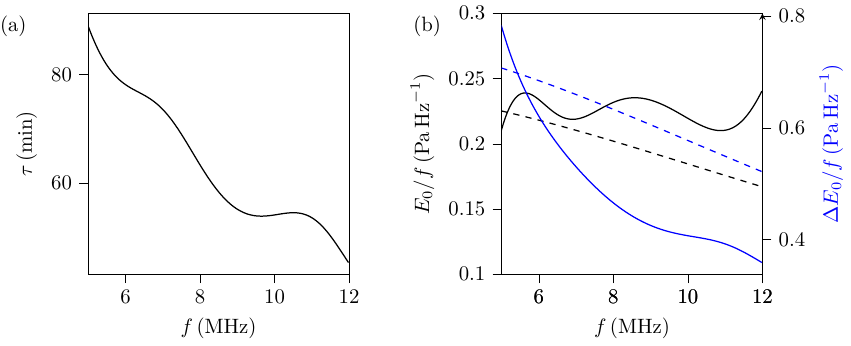}
\caption{Evolution with frequency of the parameters of the exponential fit of the viscous acoustic modulus along Eq.~\eqref{eq:exponential_fit_modulus_Esec}, for the sample of volume fraction $\varphi=\SI{4.9}{\percent}$ and salt concentration $[\mathrm{NaCl}]=\SI{0.55}{\mole\per\liter}$. (a) Characteristic time $\tau$. (b) Initial value $E_0$ (in black, left axis) and variation amplitude $\Delta E$ (in blue, right axis) divided by frequency $f$ (see discussion in section~\ref{subsec4_3}). Dashed lines correspond to the predictions of Eq.~\eqref{eq:correspondance_module_relaxation} with $\rho = \SI{e3}{\kilo\gram\per\meter\cubed}$, $c=\SI{1520}{\meter\per\second}$, $\theta=\SI{5.7e-8}{\second}$, $B_0=\SI{3.4e-2}{Np\per\meter\per\mega\hertz\squared}$ and $\Delta B=\SI{0.11}{Np\per\meter\per\mega\hertz\squared}$.}
\label{fig:US_spectra_modulus_expfit}
\end{figure*}

\begin{figure}[!htb]
\centering
\includegraphics{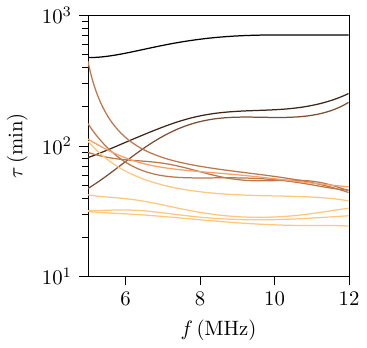}
\caption{Evolution of characteristic time of evolution of the viscous acoustic modulus (defined by Eq.~\eqref{eq:exponential_fit_modulus_Esec}) with frequency in semilogarithmic scale, for gels with identical particle volume fraction $\varphi = \SI{5.9}{\percent}$ and varying salt concentration $[\mathrm{NaCl}]=0.35$, $0.4$, $0.45$, $0.50$, $0.55$ and $\SI{0.60}{\mole\per\liter}$. Lighter colors correspond to more concentrated brines.}\label{fig:US_spectra_modulus_expfit_tauvsfreq}
\end{figure}

The evolution of these parameters with frequency are plotted on Fig.~\ref{fig:US_spectra_modulus_expfit}. We keep the discussion of parameters $E_0$ and $\Delta E$ for Section~\ref{subsec4_3}. We observe that the characteristic time $\tau$ decreases with frequency. The results for the different samples, with varying brine concentration, are gathered on Fig.~\ref{fig:US_spectra_modulus_expfit_tauvsfreq}. It is uneasy to draw general conclusions from these measurements. In agreement with the results on averaged attenuation, we observe that more concentrated brines lead to faster gelation at most frequencies, despite some sample to sample variability. Some inversions can be observed below $\SI{7}{\mega\hertz}$ but this corresponds to frequencies where signal to noise ratio is not very large and we do not consider these as significant. For the least concentrated samples, $\tau$ shows a trend to increase with frequency, while it decreases for the most concentrated samples: this calls for a more systematic investigation. However, in all cases (except for the most concentrated samples), variations with frequency are not negligible while they were hidden in the averaged approach.

\begin{figure}[!htb]
\centering
\includegraphics{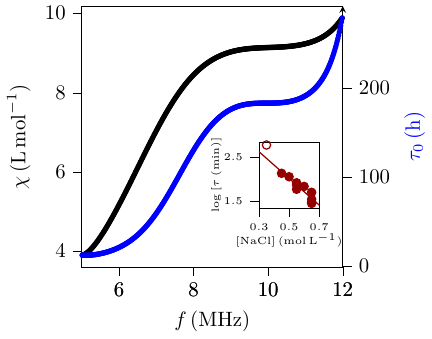}
\caption{Evolution with frequency of the parameters $\chi$ (left axis, in black) and $\tau_0$ (right axis, in blue) of the exponential fit~\eqref{eq:tau_exp_conc_frequency}, for the variations of characteristic time of viscous acoustic modulus with salt concentration. Typical uncertainty for parameter $\chi$ is the same at all frequencies, of about $\SI{0.5}{\liter\per\mole}$. Inset: Example of evolution of the characteristic time $\tau$ with salt concentration $[\mathrm{NaCl}]$, at a frequency $f=\SI{7}{\mega\hertz}$. The line displays an exponential fit according to Eq.~\eqref{eq:tau_exp_conc_frequency} with $\tau_0 = \SI{54}{\hour}$ and $\chi = \SI{6.9(5)}{\liter\per\mole}$. The open symbol corresponds to the sample at $[\mathrm{NaCl}]=\SI{0.35}{\mole\per\liter}$ that did not fully gelify over the experiment duration and is excluded from the fit.}\label{fig:US_spectra_modulus_expfit_tauvsconc}
\end{figure}

We observe that at every frequency, the characteristic time still evolves exponentially with concentration, following a generalization of Eq.~\eqref{eq:tau_exp_conc} with:
\begin{equation}
\tau(f,[\mathrm{NaCl}]) = \tau_0(f) e^{-\chi(f) [\mathrm{NaCl}]}.
\label{eq:tau_exp_conc_frequency}
\end{equation}
\noindent The inset of Fig.~\ref{fig:US_spectra_modulus_expfit_tauvsconc} shows an example of exponential fit. Again, data corresponding at brine concentration $[\mathrm{NaCl}]$ which did not fully gelify are displayed but excluded from fits. The evolution of parameters $\tau_0$ and $\chi$ are plotted on Fig.~\ref{fig:US_spectra_modulus_expfit_tauvsconc}.

We observe that the characteristic inverse concentration $\chi$ increases with frequency while characteristic time $\tau_0$ decreases. The value of $\chi$ is comparable to this of parameter $\kappa_\text{US}$ obtained with averaged attenuation. The opposite variations of $\chi$ and $\tau_0$ is associated to a peculiarity observable on Fig.~\ref{fig:US_spectra_modulus_expfit_tauvsfreq}: at low brine concentration, the gelation time $\tau$ tends to increase with frequency, while it decreases at high brine concentration. For a concentration around $[\mathrm{NaCl}] \simeq \SI{0.55}{\mole\per\liter}$, there is only little effect of frequency on the gelation time.

\subsection{Relaxation model}
\label{subsec4_2}

By proposing an exponential fit to describe data, the previous approach is inline with the analysis of average attenuation and allows to characterize gelation dynamics at different frequencies. It however remains empirical and it is interesting to attempt to model more precisely the acoustic spectra. Defining a precise model for acoustic propagation in colloidal gels is out of the scope of our present work, however one can follow a more phenomenological path.

Sound attenuation in a viscoelastic medium can be shown to be quadratic in frequency following~\cite{szabo_2000,dukhin_2010}:
\begin{equation}
\alpha = \frac{\eta_\text{tot} f^2}{2 \rho c^3} = \frac{\eta_\text{tot}}{2} \sqrt{\frac{\rho}{E^3}} f^2
\end{equation}
\noindent where $\rho$ is the density of the material, $E$ is the elastic compressibility modulus, $c=\sqrt{E/\rho}$ the speed of sound and $\eta_\text{tot} = \eta_\text{b} + 4\eta / 3$ corresponds to viscous dissipation, due to both regular shear viscosity $\eta$ and second viscosity $\eta_\text{b}$ associated with compressibility effects. We thus focus on the rescaled attenuation $\varphi(f) = \alpha/f^2$

\begin{figure*}[!htb]
\centering
\includegraphics{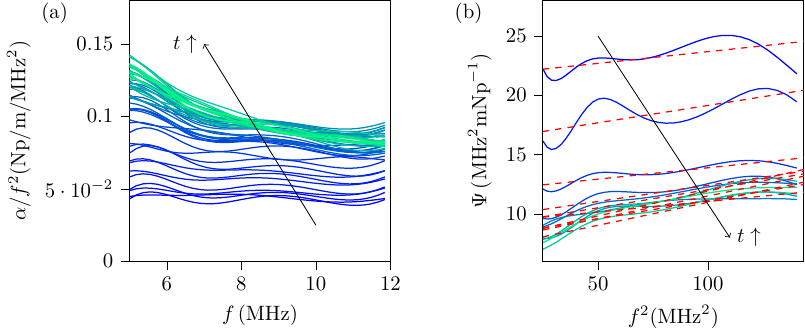}
\caption{Analysis of attenuation spectra for the sample of volume fraction $\varphi=\SI{4.9}{\percent}$ and salt concentration $[\mathrm{NaCl}]=\SI{0.55}{\mole\per\liter}$. (a) Ratio $\alpha/f^2$, for different times regularly spaced by $\SI{6}{\minute}$, lighter colors corresponding to longer times. For a newtonian liquid, this should be independent on frequency (for pure water in this frequency range, $\alpha/f^2  =\SI{5.47e-4}{Np\per\meter\per\mega\hertz\squared}$). (b) Evolution of parameter $\Psi$ defined in Eq.~\eqref{eq:definition_parameter_psi} with squared frequency and affine regression (red dashed lines). Different curves are separated by a constant time $\SI{30}{\minute}$ with bright colors indicating longer times.}
\label{fig:US_spectra_relaxation_example_tot}
\end{figure*}

An example of the variations of $\varphi$ with frequency is plotted on Fig.~\ref{fig:US_spectra_relaxation_example_tot}(a). While this rescaling dampens a significant part of the variations, we still keep a trend of decreasing $\varphi$. Such variations could be due to relaxation phenomena resulting from the interaction between the acoustic wave and the rigid skeleton of the gel. If we assume this relaxation is governed by a single time scale $\theta$, the attenuation should become~\cite{markham_1951,bryant_1999,shilov_2002,jimenez_2016}:
\begin{equation}
\alpha = A f^2 + B \frac{f^2}{1+(f\theta)^2}
\label{eq:attenuation_relaxation_model}
\end{equation}
\noindent which corresponds to a lorentzian profile of $\varphi(f)$.

A direct fit of attenuation $\alpha$ (or rescaled attenuation $\varphi$) with Eq.~\eqref{eq:attenuation_relaxation_model} was not numerically stable. In order to simplify the analysis, we decided to set the value of $A$ to the reference value $A_0=\SI{5.47e-16}{Np\per\meter}$ for water~\cite{markham_1951,holmes_2011}. We can then adjust as an affine function of $f^2$ the parameter $\Psi$ defined by:
\begin{equation}
\Psi= \frac{1}{ (\alpha/f^2) - A_0} = \frac{1}{B} \left[ 1 + (\theta f)^2 \right]
\label{eq:definition_parameter_psi}
\end{equation}
\noindent Examples of such fit for a given sample are plotted on Fig.~\ref{fig:US_spectra_relaxation_example_tot}(b). It is interesting to note that parameter $\Psi$ is actually related to the excess attenuation with respect to water which is directly measured in our experiment.

\begin{figure*}[!htb]
\centering
\includegraphics{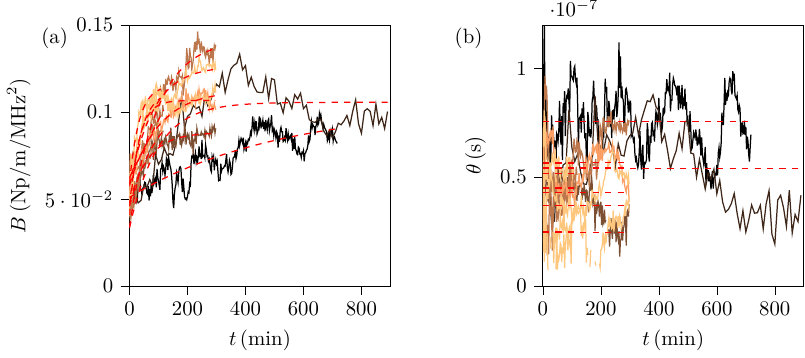}
\caption{Evolution in time of the parameters $B$ and $\theta$ defined in Eq.~\eqref{eq:attenuation_relaxation_model} for gels with identical particle volume fraction $\varphi = \SI{5.9}{\percent}$ and varying salt concentration $[\mathrm{NaCl}]=0.35$, $0.4$, $0.45$, $0.50$, $0.55$ and $\SI{0.60}{\mole\per\liter}$. Lighter colors correspond to more concentrated brines. (a) Parameter $B$. Red dashed line indicate fit along Eq.~\ref{eq:exponential_model_parameter_B}. (b) Parameter $\theta$. Red dashed line indicate average value in time $\langle \theta \rangle$. } \label{fig:US_spectra_relaxation_parameter_tot}
\end{figure*}

\begin{figure}[!htb]
\centering
\includegraphics{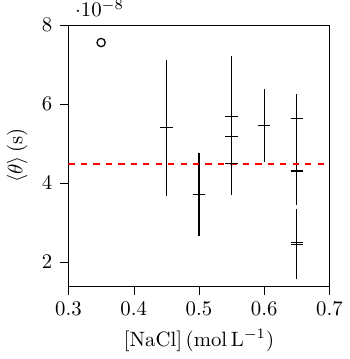}
\caption{Evolution of the average relaxation time $\langle \theta \rangle$ with the salt concentration. The dashed line indicates the average value $\theta_\text{avg} = \SI{4e-8}{\second}$, excluding the sample at $[\mathrm{NaCl}]=\SI{0.35}{\mole\per\liter}$ that did not fully gelify over the experiment duration (open symbol). Errorbars are given by the standard deviation of the time evolution of $\theta(t)$.}\label{fig:US_spectra_relaxation_thetaavg_vs_conc}
\end{figure}

By fitting spectra during gelation, we obtained time-dependent parameters $B$ and $\theta$. The evolution of the relaxation timescale $\theta$ is represented as a function of time for the different brine concentrations on Fig.~\ref{fig:US_spectra_relaxation_parameter_tot}(b). We observe that despite some important fluctuations, its evolution shows no significant trend in time. The average relaxation time $\theta$ also shows no clear trend with the brine concentration, as illustrated on Fig.~\ref{fig:US_spectra_relaxation_thetaavg_vs_conc}. This suggests that $\theta$ is associated to local gel structure rather than being a general feature of the material.

\begin{figure*}[!htb]
\centering
\includegraphics{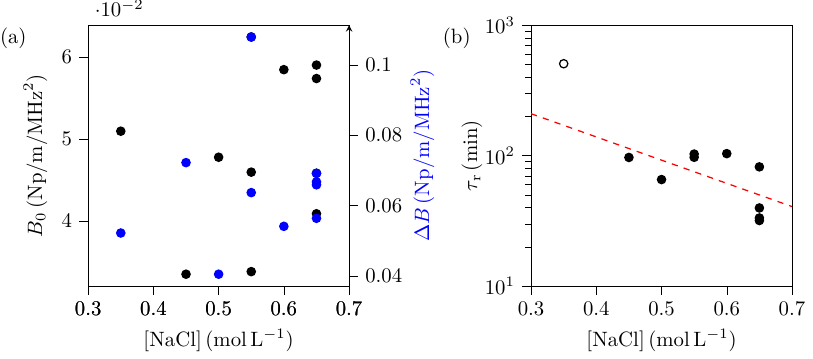}
\caption{Evolution with brine concentration of the parameters defined in Eq.~\eqref{eq:exponential_model_parameter_B}. (a) Initial value $B_0$ (black symbols, left axis) and amplitude of variation $\Delta B$ (blue symbols, right axis). (b) Characteristic time $\tau_\text{r}$. The red dashed line corresponds to an exponential fit along Eq.~\eqref{eq:tau_r_vs_conc_exp}, with $\tau_{0,\text{r}} = \SI{7.2e2}{\minute}$ and $\chi_\text{r} = \SI{4.1(5)}{\liter\per\mole}$.}
\label{fig:US_spectra_relaxation_fit_exp_tot}
\end{figure*}

Conversely, as can be seen on Fig.~\ref{fig:US_spectra_relaxation_parameter_tot}(a), parameter $B$ which quantifies the amplitude of the relaxation increases significantly in time and can be modeled by an exponential along:
\begin{equation}
B(t) = B_0 + \Delta B \left(1 - \mathrm{e}^{-t/\tau_\text{r}} \right).
\label{eq:exponential_model_parameter_B}
\end{equation}
\noindent Such a fit allows to define for all gel an initial value $B_0$ and an amplitude variation $\Delta B$ as well as a characteristic time $\tau_\text{r}$, that are plotted versus brine concentration on Fig.~\ref{fig:US_spectra_relaxation_fit_exp_tot}.

As can be seen on Fig.~\ref{fig:US_spectra_relaxation_fit_exp_tot}(a), parameters $B_0$ and $\Delta B$ show no specific trend with brine concentration and seem to be rather random parameters, specific to a given gel. The timescale $\tau_\text{r}$ that quantifies the gelation dynamics has a trend of decreasing with brine concentration, as pictured on Fig.~\ref{fig:US_spectra_relaxation_fit_exp_tot}(b). Even if the result is less convincing than with the previously defined times, one can still adjust these variations with an exponential along:
\begin{equation}
\tau_\text{r} = \tau_{0,\text{r}} e^{-\chi_\text{r} [\mathrm{NaCl}]}
\label{eq:tau_r_vs_conc_exp}
\end{equation}
\noindent We obtain a typical inverse concentration $\chi_\text{r} = \SI{4.1(5)}{\liter\per\mole}$ which is notably smaller than the previous values, but remains of the same order of magnitude.

Other models for attenuation spectrum than the proposed relaxation ~\eqref{eq:attenuation_relaxation_model} exist. For instance, a powerlaw attenuation $\alpha = \alpha_0 f^\gamma$ corresponding to a fractional Kelvin-Voigt model~\cite{holm_2010,holm_2011}) also provides us with an acceptable description of the spectra, but the evolution in time of the resulting parameters $\alpha_0$ and $\gamma$ show a poor reproducibility. At this stage, it is thus not possible to draw clear conclusions about the acoustic spectra. These results therefore remain preliminary and improved experimental data would be helpful, in particular with an extended frequency range to better constrain fits.

\subsection{Comparison of the two approaches}
\label{subsec4_3}

It is finally interesting to compare the two approaches presented in Sections~\ref{subsec4_1} and~\ref{subsec4_2}. By injecting the expression of $\alpha$ given by the relaxation model~\eqref{eq:attenuation_relaxation_model} in the expression of viscous acoustic modulus~\eqref{eq:definition_modulus_Esec}, and considering the limit $\alpha c \ll f$, we obtain:
\begin{equation}
E'' = 2 \rho c^3 \left[ A_0 + \frac{B(t)}{1+ (f\theta)^2} \right] f .
\end{equation}
\noindent Using the exponential evolution obtained for $B(t)$ in Eq.~\eqref{eq:exponential_model_parameter_B}, we finally obtain:
\begin{equation}
E'' = 2 \rho c^3 \left[ A_0 + \frac{B_0}{1+ (f\theta)^2} \right] f + 2\rho c^3 \frac{\Delta B}{1+ (f\theta)^2} f \left(1- \mathrm{e}^{-t/\tau_\text{r}} \right).
\end{equation}
\noindent This is compatible with the exponential evolution of $E''$ given by Eq.~\eqref{eq:exponential_fit_modulus_Esec}, provided that $\tau = \tau_\text{r}$ and:
\begin{equation}
\frac{E_0}{f} = 2 \rho c^3 \left[ A_0 + \frac{B_0}{1+ (f\theta)^2} \right] \; \text{and} \; \frac{\Delta E}{f} = 2\rho c^3 \frac{\Delta B}{1+ (f\theta)^2}.
\label{eq:correspondance_module_relaxation}
\end{equation}

In order to discuss the validity of these conditions, we consider the sample of brine concentration $[\mathrm{NaCl}] = \SI{0.55}{\mole\per\liter}$ which has been used as an example throughout the entire note but the main conclusions are similar for all samples. The prediction of Eq.~\eqref{eq:correspondance_module_relaxation} are plotted as dashed line on Fig.~\ref{fig:US_spectra_modulus_expfit}(b), by taking $\rho=\SI{e3}{\kilo\gram\per\meter\cubed}$ (corresponding to the density of the solution obtained by mixing brine and Ludox suspension), $c=\SI{1520}{\meter\per\second}$ (corresponding to the average sound speed) and values of $\theta$, $B_0$ and $\Delta B$ are taking as averages over frequency. While the agreement is not fully satisfactory, we still obtain acceptable trends and orders of magnitude. For this sample, $\tau_\text{r} = \SI{97}{\minute}$ : again, the order of magnitude is comparable to this obtained in Fig.~\ref{fig:US_spectra_modulus_expfit}(a).

However, this discussion assumes that the different parameters are frequency-independent: while this is not incorrect for $c$, $B_0$ and $\Delta B$, this is far more debatable for $\tau$. With our current results, it is not possible to push further this discussion: the comparison between the two approaches shows no clear incompatibility, but a more in-depth confrontation would require data of better quality (in particular on an extended frequency range) and reproducing the experiment to improve statistics.



\section{Conclusion and perspectives}
\label{sec5}

In this note, we used a simple setup of ultrasonic spectroscopy to monitor the gelation process of a colloidal gel. We more precisely studied the gelation of colloidal silica due to the mixing with a concentrated brine, which is a model system for physical gelation but had not been studied with acoustic techniques previously. In this system, it has been shown in the literature that the brine concentration allows to control the gelation dynamics. We showed that while sound speed remains almost constant, essentially corresponding to its value in water, acoustic attenuation increases in time.

In a first step, we only considered average attenuation. It allowed us to define a gelation time, that compares satisfactorily with common rheological measurements. This shows the interest of acoustic spectroscopy for monitoring of dynamical processes, as such a setup can easily be adapted to industrial contexts. Moreover, we showed that our setup is capable of mapping spatially the acoustic properties of the gel, and reveals heterogeneities of the gel structure.

Then, we proposed a more careful study of the acoustic spectra. Our measurement first show that overall, acoustic attenuation increases in time at all frequencies and with increasing brine concentration. As a first approach, fitting acoustic moduli as a function of time, frequency per frequency, allows to characterize the frequency-dependent gelation dynamics, while remaining in agreement with the overall trends observed with the average approach. Then, we used a more physical description of acoustic spectra using a single timescale relaxation model, that also allows to characterize the gelation dynamics. These two points of view can be compared: some incompatibilities exist with our current data, but lack of statistics makes difficult to clearly decide the best approach. Our work however gives a good description of attenuation spectra of physical colloidal gels.

The current study would require more experiments to perform more precise statistics, and an improved setup to access a wider frequency range. It however shows that acoustic spectroscopy offers opportunities for the study of physico-chemical transformations in soft materials, both for applied perspective of contactless and non-intrusive monitoring, and more fundamental characterization of sound propagation in complex media. We hope it will generate new efforts at the interface between acoustics and rheology, in particular to clarify the connection between acous

\section*{Acknowledgments}

We thank the R{\'e}gion Nouvelle-Aquitaine for funding through the Holofluid-3D project and Solvay for funding NB PhD. We are grateful to Dr. Thomas Brunet for interesting discussions about our setup and results. We also thank Pr. S{\'e}bastien Manneville and Dr. Thomas Gibaud for suggestions on Ludox gels formulation and Dr. Valentin Leroy for bringing interesting references to our knowledge. We are finally grateful to an anonymous referee of a first version of this article for suggesting relevant analysis of our data that led to sections~\ref{subsec4_2} and~\ref{subsec4_3}.

\section*{Declarations}

\begin{itemize}
\item Funding: R{\'e}gion Nouvelle-Aquitaine funded material acquisition for the present work through the Holofluid-3D project and Solvay funded NB PhD.
\item Employment: Not applicable.
\item Financial interests: Not applicable.
\item Non financial interests: Not applicable.
\item Availability of data: The datasets generated during and/or analysed during the current study are available from the corresponding author on reasonable request.
\item Authors' contributions: JNT and PL conceived the project and obtained fundings. NB, JNT and PL conceived the experiments and supervised the project. NB, MJNB and JNT conducted the experiments. NB and PL analyzed data and wrote the manuscript. NB, JNT and PL reviewed the manuscript.
\end{itemize}

\appendix

\section{Details on acoustic signal analysis}
\label{A_sec1}

In this appendix, we provide the reader with more details on the treatment of acoustic signal to retrieve quantities of interest. All signal analysis is coded in Python and uses libraries Numpy, SciPy and Pandas.

\subsection{Analysis of a acoustic signal}

\begin{figure*}[!htb]
\centering
\includegraphics[width=0.9\textwidth]{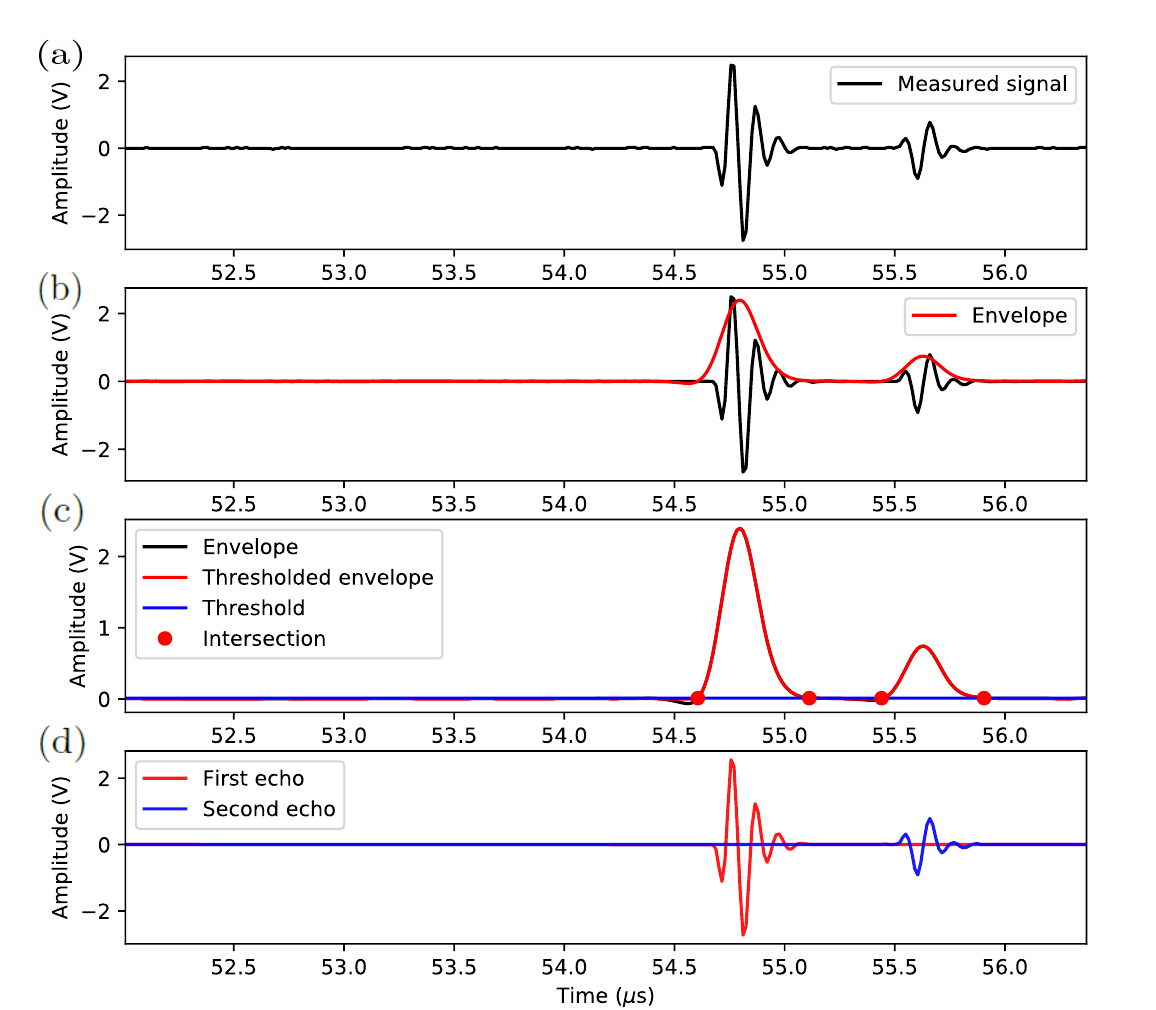}
\caption{Illustration of the method used to isolate echoes on a signal, for a measurement in reflection mode. The initial signal (a) shows two echoes corresponding to the reflections at the two interfaces of the window. The envelope of the signal is extracted by taking its analytical part. The obtained envelope is then thresholded to retrieve the times at which a given echo starts and stops (c). The isolated echoes are highlighted in different colors (d).}\label{fig:appendix_isolate_signal}
\end{figure*}

Figure~\ref{fig:appendix_isolate_signal} shows the different steps to isolate a single peak on a given signal. Figure~\ref{fig:appendix_isolate_signal}(a) presents an example of signal observed in transmission, with two successive pulses corresponding to different trajectories in the sample: in order to further analyze the signal, it is necessary to separate these echoes. The cell is built such that there is no overlap between the different echoes, which simplifies the procedure. It is to note that the signal is actually averaged over $10$ successive excitations of the emitting transducer in order to improve the signal to noise ratio. We then extract the envelope of the signal by taking its analytical part, using Hilbert transform (Fig.~\ref{fig:appendix_isolate_signal}(b)): by applying a simple threshold, we finally determine the support of the different echoes (Fig.~\ref{fig:appendix_isolate_signal}(c)) which allows to separate them from each other.

\subsection{Calibration procedure and measurement of sound speed and attenuation}

When the different echoes of the signals measured on the two transducers are isolated, we compute their Fourier transforms. In order to measure the excess time of flight and attenuation of the sample with respect to water, we need to consider:
\begin{itemize}
\item the signal $u_\text{a,r}$ obtained after a single reflection on the cell filled with air,
\item the signal $u_\text{w,r}$ obtained after a single reflection on the cell filled with water,
\item the signal $u_\text{w,t}$ obtained after direct transmission through the cell filled with water,
\item the signal $u_\text{s,r}$ obtained after a single reflection on the cell filled with the sample,
\item the signal $u_\text{s,t}$ obtained after direct transmission through the cell filled with the sample. 
\end{itemize}
\noindent We denote by $u^0$ the spectral content of the acoustic wave initially generated by the emitting transducer. All the quantities in this section depend on frequency but we do not write it for the sake of simplicity. We denote by $r_{1/2}$ and $t_{1/2} = 1+r_{1/2}$ the reflection and transmission coefficient in amplitude for a wave arriving on an interface between two media $(1)$ and $(2)$ from medium $(1)$, and we have $r_{1/2}=-r_{2/1}$~\cite{kinsler}. Here, the different possible media are water (w), air (a), polycarbonate window (PC) and the sample under study (s). We finally denote by $\ell$ the cell thickness.

By writing the phase and attenuation due to propagation and the coefficients associated with reflection and transmission through interfaces for the different signals, and by assuming that the interface between the polycarbonate window and air is an ideal reflector ($r_\text{PC/a}=1$), we can calibrate the experiment and retrieve the quantities we are interested in. First, by considering the ratio between the reflection on air and on the sample, and on air and on water, we respectively obtain the reflection coefficients on the window:
\begin{equation}
r_\text{w/PC} = \frac{u_\text{w,r}}{u_\text{a,r}} \text{ and } r_\text{s/PC} = \frac{u_\text{s,r}}{u_\text{a,r}}.
\end{equation}
\noindent From these coefficients, we can also easily deduce the associated transmission coefficients. Knowing these factors, the excess attenuation and time of flight in sample with respect to water can obtained by normalizing the transmitted signal through the sample and through the water:
\begin{equation}
\alpha_\text{s} - \alpha_\text{w} = \frac{1}{\ell} \ln \Bigg \lvert \frac{1-r_\text{s/PC}^2}{1-r_\text{w/PC}^2} \frac{u_\text{w,t}}{u_\text{s,t}} \Bigg \rvert
\end{equation}
\noindent and:
\begin{equation}
\frac{\ell}{c_\text{s}} - \frac{\ell}{c_\text{w}} = \frac{1}{2\pi f \mathrm{arg}\left[ \frac{1-r_\text{s/PC}^2}{1-r_\text{w/PC}^2}\right]}\left[ \mathrm{arg}\left(\frac{u_\text{s,t}}{u_\text{w,t}}\right) + 2q\pi \right]
\end{equation}
\noindent where $q$ is an integer. This parameter is due to the fact that the phase is known modulo $2\pi$. In order to measure it, we unwrap the phase of $u_\text{s,t}/u_\text{w,t}$ which is almost linear with frequency, and extrapolate it at null frequency, giving a value $-A$. The physical phase difference between the two signals should cancel in this limit, so we take $q$ as the floor of $A$.

Attenuation $\alpha_\text{w}$ and sound speed $c_\text{w}$ in water are taken from reference values~\cite{holmes_2011,marczak_1997}. Finally, $\ell$ is obtained by measuring the difference of time of flight for a cell filled with water between the reflection at the first window/water interface (corresponding to signal $u_\text{w,r}$) and at the water/second window interface.


\begin{thebibliography}{10}
\providecommand{\url}[1]{{#1}}
\providecommand{\urlprefix}{URL }
\providecommand{\doi}[1]{\url{https://doi.org/#1}}

\bibitem{bercoff_2004}
J.~Bercoff, M.~Tanter, M.~Fink, Supersonic shear imaging: A new technique for
  soft tissue elasticity mapping.
\newblock IEEE transactions on ultrasonics, ferroelectrics, and frequency
  control \textbf{51}, 396--409 (2004)

\bibitem{errico_2015}
C.~Errico, J.~Pierre, S.~Pezet, Y.~Desailly, Z.~Lenkei, O.~Couture, M.~Tanter,
  Ultrafast ultrasound localization microscopy for deep super-resolution
  vascular imaging.
\newblock Nature \textbf{527}, 499--507 (2015).
\newblock \doi{10.1038/nature16066}

\bibitem{wiklund_2007}
J.~Wiklund, I.~Shahram, M.~Stading, Methodology for in-line rheology by
  ultrasound doppler velocity profiling and pressure difference techniques.
\newblock Chemical Engineering Science \textbf{62}, 4277--4293 (2007).
\newblock \doi{10.1016/j.ces.2007.05.007}

\bibitem{gallot_2013}
T.~Gallot, C.~Perge, V.~Grenard, M.~Fardin, N.~Taberlet, S.~Manneville,
  Ultrafast ultrasonic imaging coupled to rheometry: Principle and
  illustration.
\newblock Review of Scientific Instruments \textbf{84}, 045,107 (2013).
\newblock \doi{10.1063/1.4801462}

\bibitem{mograne_2019}
M.~Mograne, J.~Ferrandis, D.~Laux, Instrumented test tube for rapid rheological
  behaviour of liquids estimation.
\newblock Journal of Food Engineering \textbf{247}, 126--129 (2019).
\newblock \doi{10.1016/j.jfoodeng.2018.12.007}

\bibitem{markham_1951}
J.~Markham, R.~Beyer, R.~Lindsay, Absorption of sound in fluids.
\newblock Reviews of Modern Physics \textbf{23}, 353--411 (1951)

\bibitem{hirai_1958}
N.~Hirai, H.~Eyring, Bulk viscosity of liquids.
\newblock Journal of Applied Physics \textbf{29}, 810--816 (1958).
\newblock \doi{10.1063/1.1723290}

\bibitem{eggers_1996}
F.~Eggers, U.~Kaatze, Broad-band ultrasonic measurement techniques for liquids.
\newblock Measurement Science and Technology \textbf{7}, 1--19 (1996)

\bibitem{kaatze_2000}
U.~Kaatze, T.~Hushcha, F.~Eggers, Broad-band ultrasonic measurement techniques
  for liquids.
\newblock Journal of Solution Chemistry \textbf{29}, 299--368 (2000)

\bibitem{longin_1998}
P.~Longin, C.~Verdier, M.~Piau, Dynamic shear rheology of high molecular weight
  polydimethylsiloxanes: comparison of rheometry and ultrasound.
\newblock Journal of Non-Newtonian Fluid Mechanics \textbf{76}, 213--232 (1998)

\bibitem{verdier_1998}
C.~Verdier, P.~Longin, M.~Piau, Dynamic shear and compressional behavior of
  polydimethylsiloxanes: Ultrasonic and low frequency characterization.
\newblock Rheologica Acta \textbf{37}, 234--244 (1998)

\bibitem{leroy_2010}
V.~Leroy, K.~Pitura, M.~Scanlon, J.~Page, The complex shear modulus of dough
  over a wide frequency range.
\newblock Journal of Non-Newtonian Fluid Mechanics \textbf{165}, 475--478
  (2010).
\newblock \doi{10.1016/j.jnnfm.2010.02.001}

\bibitem{scanlon_2015}
M.~Scanlon, J.~Page, Probing the properties of dough with low-intensity
  ultrasound.
\newblock Cereal Chemistry \textbf{92}, 121--133 (2015).
\newblock \doi{10.1094/CCHEM-11-13-0244-IA}

\bibitem{lefebvre_2018}
G.~Lefebvre, R.~Wunenburger, T.~Valier-Brasier, Ultrasonic rheology of
  visco-elastic materials using shear and longitudinal waves.
\newblock Applied Physics Letters \textbf{112}, 241,906 (2018).
\newblock \doi{10.1063/1.5029905}

\bibitem{forrester_2016a}
D.~Forrester, J.~Huang, V.~Pinfield, Characterisation of colloidal dispersions
  using ultrasound spectroscopy and multiple-scattering theory inclusive of
  shear-wave effects.
\newblock Chemical Engineering Research and Design \textbf{114}, 69--78 (2016).
\newblock \doi{10.1016/j.cherd.2016.08.008}

\bibitem{forrester_2016b}
D.~Forrester, J.~Huang, V.~Pinfield, F.~Lupp{\'e}, Experimental verification of
  nanofluid shear-wave reconversion in ultrasonic fields.
\newblock Nanoscale \textbf{8}, 84--88 (2016).
\newblock \doi{10.1039/c5nr07396k}

\bibitem{mori_2018}
H.~Mori, T.~Norisuye, H.~Nakanishi, Q.~Tran-Cong-Miyata, Ultrasound attenuation
  and phase velocity of micrometer-sized particlesuspensions with viscous and
  thermal losses.
\newblock Ultrasonics \textbf{83}, 171--178 (2018).
\newblock \doi{10.1016/j.ultras.2017.03.016}

\bibitem{epstein_1953}
P.~Epstein, R.~Carhart, The absorption of sound in suspensions and emulsions.
  i. water fog in air.
\newblock Journal of the Acoustical Society of America \textbf{25}, 553--565
  (1953)

\bibitem{allegra_1972}
J.~Allegra, S.~Hawley, Attenuation of sound in suspensions and emulsions:
  Theory and experiments.
\newblock Journal of the Acoustical Society of America \textbf{51}, 1545--1564
  (1972)

\bibitem{challis_2005}
R.~Challis, M.~Povey, M.~Mather, A.~Holmes, Ultrasound techniques for
  characterizing colloidal dispersions.
\newblock Reports on Progress in Physics \textbf{68}, 1541--1637 (2005).
\newblock \doi{10.1088/0034-4885/68/7/R01}

\bibitem{allashi_2014}
R.~Al-Lashi, R.~Challis, Effective viscosity in a wave propagation model for
  ultrasonic particle sizing in non-dilute suspensions.
\newblock Journal of the Acoustical Society of America \textbf{136}, 1583--1890
  (2014).
\newblock \doi{10.1121/1.4894689}

\bibitem{pinfield_2015}
V.~Pinfield, D.~Forrester, F.~Lupp{\'e}, Ultrasound propagation in concentrated
  suspensions: shear-mediated contributions to multiple scattering.
\newblock Physics Procedia \textbf{70}, 213--216 (2015).
\newblock \doi{10.1016/j.phpro.2015.08.135}

\bibitem{valierbrasier_2015}
T.~Valier-Brasier, J.~Conoir, F.~Coulouvrat, J.~Thomas, Sound propagation in
  dilute suspensions of spheres: Analytical comparison between coupled phase
  model and multiple scattering theory.
\newblock Journal of the Acoustical Society of America \textbf{138}, 2598--2612
  (2015).
\newblock \doi{10.1121/1.4932171}

\bibitem{forrester_2019}
D.~Forrester, J.~Huang, V.~Pinfield, Modelling viscous boundary layer
  dissipation effects in liquid surrounding individual solid nano and
  micro-particles in an ultrasonic field.
\newblock Scientific Reports \textbf{9}, 4956 (2019).
\newblock \doi{10.1038/s41598-019-40665-9}

\bibitem{dukhin_1996b}
A.~Dukhin, P.~Goetz, Acoustic spectroscopy for concentrated polydisperse
  colloids with high density contrast.
\newblock Langmuir \textbf{12}, 4987--4997 (1996)

\bibitem{dukhin_characterization}
A.~Dukhin, P.~Goetz, \emph{Characterization of Liquids, Nano- and
  Microparticulates, and Porous Bodies using Ultrasound} (Elsevier, 2010)

\bibitem{maleky_2007}
F.~Maleky, R.~Campos, A.~Marangoni, Structural and mechanical properties of
  fats quantified by ultrasonics.
\newblock Journal of the American Oil Chemists Society \textbf{84}, 331--338
  (2007).
\newblock \doi{10.1007/s11746-007-1039-3}

\bibitem{maleky_2011}
F.~Maleky, A.~Marangoni, Ultrasonic technique for determination of the shear
  elastic modulus of polycrystalline soft materials.
\newblock Crystal Growth and Design \textbf{11}, 941--944 (2011).
\newblock \doi{10.1021/cg2000188}

\bibitem{franco_2019}
E.~Franco, F.~Buiochi, Ultrasonic measurement of viscosity: Signal processing
  methodologies.
\newblock Ultrasonics \textbf{91}, 213--219 (2019).
\newblock \doi{10.1016/j.ultras.2018.08.006}

\bibitem{giordano_1983}
R.~Giordano, F.~Mallamace, F.~Wanderlingh, Acoustic absorption and thixotropic
  structure in lysozymc solution.
\newblock Il Nuovo Cimento \textbf{2D}, 1272--1280 (1983)

\bibitem{dalgleish_2005}
D.~Dalgleish, E.~Verespej, M.~Alexander, M.~Corredig, The ultrasonic properties
  of skim milk related to the release of calcium from casein micelles during
  acidification.
\newblock International Dairy Journal \textbf{15}, 1105--1112 (2005).
\newblock \doi{10.1016/j.idairyj.2004.11.016}

\bibitem{kuo_2008}
F.~Kuo, C.~Sheng, C.~Ting, Evaluation of ultrasonic propagation to measure
  sugar content and viscosity of reconstituted orange juice.
\newblock Journal of Food Engineering \textbf{86}, 84--90 (2008).
\newblock \doi{10.1016/j.jfoodeng.2007.09.016}

\bibitem{resa_2009}
P.~Resa, L.~Elvira, F.~Montero~de Espinosa, R.~Gonz{\'a}lez, J.~Barcenilla,
  On-line ultrasonic velocity monitoring of alcoholic fermentation kinetics.
\newblock Bioprocess and Biosystems Engineering \textbf{32}, 321--331 (2009).
\newblock \doi{10.1007/s00449-008-0251-3}

\bibitem{laux_2013}
D.~Laux, M.~Valente, J.~Ferrandis, N.~Talha, O.~Gibert, A.~Prades, Shear
  viscosity investigation on mango juice with high frequency longitudinal
  ultrasonic waves and rotational viscosimetry.
\newblock Food Biophysics \textbf{8}, 233--239 (2013).
\newblock \doi{10.1007/s11483-013-9291-6}

\bibitem{laux_2014}
D.~Laux, O.~Gibert, J.~Ferrandis, M.~Valente, A.~Prades, Ultrasonic evaluation
  of coconut water shear viscosity.
\newblock Journal of Food Engineering \textbf{126}, 62--64 (2014).
\newblock \doi{10.1016/j.jfoodeng.2013.11.007}

\bibitem{derra_2018}
M.~Derra, F.~Bakkali, A.~Amghar, H.~Sahsah, Estimation of coagulation time in
  cheese manufacture using an ultrasonic pulse-echo technique.
\newblock Journal of Food Engineering \textbf{216}, 65--71 (2018).
\newblock \doi{10.1016/j.jfoodeng.2017.08.003}

\bibitem{landini_1986}
L.~Landini, R.~Sarnelli, Evaluation of the attenuation coefficients in nomral
  and pathological breast tissue.
\newblock Medical \& Biological Engineering \& Computing \textbf{24}, 243--247
  (1986)

\bibitem{jongen_1986}
H.~Jongen, J.~Thijssen, M.~van~den Aarssen, W.~Verhoef, A general model for the
  absorption of ultrasound by biological tissues and experimental verification.
\newblock Journal of the Acoustical Society of America \textbf{79}, 535--540
  (1986)

\bibitem{parker_1988}
K.~Parker, M.~Asztely, R.~Lerner, E.~Schenk, R.~Waag, In-vivo measurements of
  ultrasound attenuation in normal or diseased liver.
\newblock Ultrasound in Medicine \& Biology \textbf{14}, 127--136 (1988)

\bibitem{bonacucina_2008}
G.~Bonacucina, M.~Misici-Falzi, M.~Cespi, G.~Palmieri, Characterization of
  micellar systems by the use of acoustic spectroscopy.
\newblock Journal of Pharmaceutical Sciences \textbf{97}, 2217--2227 (2008)

\bibitem{bonacucina_2016}
G.~Bonacucina, D.~Perinelli, M.~Cespi, L.~Casettari, R.~Cossi, P.~Blasi,
  G.~Palmieri, Acoustic spectroscopy: A powerful analytical method for the
  pharmaceutical field?
\newblock International Journal of Pharmaceutics \textbf{50}, 174--195 (2016).
\newblock \doi{10.1016/j.ijpharm.2016.03.009}

\bibitem{krasaekoopt_2005}
W.~Krasaekoopt, B.~Bhandari, H.~Deeth, Comparison of gelation profile of
  yoghurts during fermentation measured by rva and ultrasonic spectroscopy.
\newblock International Journal of Food Properties \textbf{8}, 193--198 (2005).
\newblock \doi{10.1081/JFP-200059469}

\bibitem{bacri_1980}
J.~Bacri, J.~Courdille, J.~Dumas, R.~Rajanoarison, Ultrasonic waves : a tool
  for gelation process measurements.
\newblock Journal de Physique Lettres \textbf{41}, 369--372 (1980).
\newblock \doi{10.1051/jphyslet:019800041015036900}

\bibitem{sidebottom_1993}
D.~Sidebottom, Ultrasonic measurements of an epoxy resin near its sol-gel
  transition.
\newblock Physical Review E \textbf{48}, 391--399 (1993)

\bibitem{parker_2012}
N.~Parker, M.~Povey, Ultrasonic study of the gelation of gelatin: Phase
  diagram, hysteresis and kinetics.
\newblock Food Hydrocolloids \textbf{26}, 99--107 (2012).
\newblock \doi{10.1016/j.foodhyd.2011.04.016}

\bibitem{serfaty_1998}
S.~Serfaty, P.~Griesmar, M.~Gindre, G.~Gouedard, P.~Figui{\`e}re, An acoustic
  technique for investigating the sol–gel transition.
\newblock Journal of Materials Chemistry \textbf{8}, 2229--2231 (1998)

\bibitem{forest_1998}
L.~Forest, V.~Gibiat, T.~Woignier, Evolution of the acoustical properties of
  silica alcogels during their formation.
\newblock Ultrasonics \textbf{36}, 477--481 (1998)

\bibitem{cros_2001}
B.~Cros, M.~Pauthe, M.~Rguiti, J.~Ferrandis, On-line characterization of silica
  gels by acoustic near field.
\newblock Sensors and Actuators B: Chemical \textbf{76}, 115--123 (2001).
\newblock \doi{10.1016/j.snb.2009.10.070}

\bibitem{griesmar_2003}
P.~Griesmar, A.~Ponton, S.~Serfaty, B.~Senouci, M.~Gindre, G.~Gouedard,
  S.~Warlus, Kinetic study of silicon alkoxides gelation by acoustic and
  rheology investigations.
\newblock Journal of Non-Crystalline Solids \textbf{319}, 57--64 (2003).
\newblock \doi{10.1016/S0022-3093(02)01957-9}

\bibitem{ouldehssein_2006}
C.~Ould~Ehssein, S.~Serfaty, P.~Griesmar, J.~Le~Huerou, L.~Martinez,
  E.~Caplain, N.~Wilkie-Chancellier, M.~Gindre, G.~Gouedard, P.~Figuiere,
  Kinetic study of silica gels by a new rheological ultrasonic investigation.
\newblock Ultrasonics \textbf{44}, 881--889 (2006).
\newblock \doi{10.1016/j.ultras.2006.05.035}

\bibitem{robin_2012}
G.~Robin, F.~Vander~Meulen, N.~Wilkie-Chancellier, L.~Martinez, L.~Haumesser,
  J.~Fortineau, P.~Griesmar, M.~Lethiecq, G.~Feuillard, Ultrasonic
  self-calibrated method applied to monitoring of sol–gel transition.
\newblock Ultrasonics \textbf{52}, 881--889 (2012).
\newblock \doi{10.1016/j.ultras.2011.12.008}

\bibitem{moller_2006}
P.~M{\o}ller, J.~Mewis, D.~Bonn, Yield stress and thixotropy: on the difficulty
  of measuring yield stresses in practice.
\newblock Soft Matter \textbf{2}, 274--283 (2006)

\bibitem{coussot_2014}
P.~Coussot, Yield stress fluid flows: A review of experimental data.
\newblock Journal of Non-Newtonian Fluid Mechanics \textbf{211}, 31--49 (2014).
\newblock \doi{10.1016/j.jnnfm.2014.05.006}

\bibitem{bonn_2017}
D.~Bonn, M.~Denn, L.~Berthier, T.~Divoux, S.~Manneville, Yield stress materials
  in soft condensed matter.
\newblock Reviews of Modern Physics \textbf{89}, 035,005 (2017).
\newblock \doi{10.1103/RevModPhys.89.035005}

\bibitem{kurokawa_2015}
A.~Kurokawa, V.~Vidal, K.~Kurita, T.~Divoux, S.~Manneville, Avalanche-like
  fluidization of a non-brownian particle gel.
\newblock Soft Matter \textbf{11}, 9026--9037 (2015).
\newblock \doi{10.1039/c5sm01259g}

\bibitem{trompette_2003}
J.~Trompette, M.~Meireles, Ion-specific effect on the gelation kinetics of
  concentrated colloidal silica suspensions.
\newblock Journal of Colloid and Interface Science \textbf{263}, 522--527
  (2003).
\newblock \doi{10.1016/S0021-9797(03)00397-7}

\bibitem{trompette_2004}
J.~Trompette, M.~Clifton, Influence of ionic specificity on the microstructure
  and the strength of gelled colloidal silica suspensions.
\newblock Journal of Colloid and Interface Science \textbf{276}, 475--482
  (2004).
\newblock \doi{10.1016/j.jcis.2004.03.040}

\bibitem{cao_2010}
X.~Cao, H.~Cummins, J.~Morris, Structural and rheological evolution of silica
  nanoparticle gels.
\newblock Soft Matter \textbf{6}, 5425--5433 (2010).
\newblock \doi{10.1039/c0sm00433b}

\bibitem{trompette_2000}
J.~Trompette, C.~Bordes, Rheological study of the gelation kinetics of a
  concentrated latex suspension in the presence of nonadsorbing polymer chains.
\newblock Langmuir \textbf{16}, 9627--9633 (2000).
\newblock \doi{10.1021/la000786e}

\bibitem{trompette_2017}
J.~Trompette, Influence of co-ion nature on the gelation kinetics of colloidal
  silica suspensions.
\newblock Journal of Physical Chemistry B \textbf{121}, 5654--5659 (2017).
\newblock \doi{10.1021/acs.jpcb.7b03007}

\bibitem{hallez_2017}
Y.~Hallez, M.~Meireles, Fast, robust evaluation of the equation of state of
  suspensions of charge-stabilized colloidal spheres.
\newblock Langmuir \textbf{33}, 10,051--10,060 (2017).
\newblock \doi{10.1021/acs.langmuir.7b02209}

\bibitem{rogers_1974}
P.~Rogers, A.~van Buren, An exact expression for the lommel diffraction
  correction integral.
\newblock Journal of the Acoustical Society of America \textbf{55}, 724--728
  (1974)

\bibitem{marczak_1997}
W.~Marczak, Water as a standard in the measurements of speed of sound in
  liquids.
\newblock Journal of the Acoustical Society of America \textbf{102}, 2776--2779
  (1997)

\bibitem{holmes_2011}
M.~Holmes, N.~Parker, M.~Povey, Temperature dependence of bulk viscosity in
  water using acoustic spectroscopy.
\newblock Journal of Physics: Conference Series \textbf{269}, 012,011 (2011).
\newblock \doi{10.1088/1742-6596/269/1/012011}

\bibitem{winter_1987}
H.~Winter, Can the gel point of a cross-linking polymer be detected by the g' -
  g'' crossover?
\newblock Polymer Engineering and Science \textbf{27}, 1698--1702 (1987)

\bibitem{chambon_1987}
F.~Chambon, H.~Winter, Linear viscoelasticity at the gel point of a
  crosslinking pdms with imbalanced stoichiometry.
\newblock Journal of Rheology \textbf{31}, 683--697 (1987)

\bibitem{holly_1988}
E.~Holly, S.~Venkataraman, F.~Chambon, H.~Winter, Fourier transform mechanical
  spectroscopy of viscoelastic materials with transient structure.
\newblock Journal of Non-Newtonian Fluid Mechanics \textbf{27}, 17--26 (1988)

\bibitem{hodgson_1990}
D.~Hodgson, E.~Amis, Dynamic viscoelastic characterization of sol-gel
  reactions.
\newblock Macromolecules \textbf{23}, 2512--2519 (1990)

\bibitem{winter_1991}
H.~Winter, Polymer gels, materials that combine liquid and solid properties.
\newblock MRS Bulletin pp. 44--48 (1991)

\bibitem{ponton_2002}
A.~Ponton, S.~Warlus, P.~Griesmar, Rheological study of the sol–gel
  transition in silica alkoxides.
\newblock Journal of Colloid and Interface Science \textbf{249}, 209--216
  (2002).
\newblock \doi{10.1016/jcis.2002.8227}

\bibitem{drabarek_2002}
E.~Drabarek, J.~Bartlett, H.~Hanley, J.~Woolfrey, C.~Muzny, Effect of
  processing variables on the structural evolution of silica gels.
\newblock International Journal of Thermophysics \textbf{23}, 145--160 (2002)

\bibitem{matsunaga_2007}
T.~Matsunaga, M.~Shibayama, Gel point determination of gelatin hydrogels by
  dynamic light scattering and rheological measurements.
\newblock Physical Review E \textbf{76}, 030,401 (2021).
\newblock \doi{10.1103/PhysRevE.76.030401}

\bibitem{zaccone_2009}
A.~Zaccone, H.~Wu, D.~Gentili, M.~Morbidelli, Theory of activated-rate
  processes under shear with application to shear-induced aggregation of
  colloids.
\newblock Physical Review E \textbf{80}, 051,404 (2009).
\newblock \doi{10.1103/PhysRevE.80.051404}

\bibitem{zaccone_2010}
A.~Zaccone, D.~Gentili, H.~Wu, M.~Morbidelli, Shear-induced reaction-limited
  aggregation kinetics of brownian particles at arbitrary concentrations.
\newblock Journal of Chemical Physics \textbf{132}, 134,903 (2010).
\newblock \doi{10.1063/1.3361665}

\bibitem{zaccone_2011}
A.~Zaccone, D.~Gentili, H.~Wu, M.~Morbidelli, E.~Del~Gado, Shear-driven
  solidification of dilute colloidal suspensions.
\newblock Physical Review Letters \textbf{106}, 138,301 (2021).
\newblock \doi{10.1103/PhysRevLett.106.138301}

\bibitem{gibaud_2020}
T.~Gibaud, N.~Dag{\`e}s, P.~Lidon, G.~Jung, L.~Ahour{\'e}, M.~Sztucki,
  A.~Poulesquen, N.~Hengl, F.~Pignon, S.~Manneville, Rheoacoustic gels: Tuning
  mechanical and flow properties of colloidal gels with ultrasonic vibrations.
\newblock Physical Review X \textbf{10}, 011,028 (2020).
\newblock \doi{10.1103/PhysRevX.10.011028}

\bibitem{dukhin_2010}
A.~Dukhin, P.~Goetz, in \emph{Studies in Interface Science, vol. 24} (Elsevier,
  2010).
\newblock \doi{10.1016/S1383-7303(10)23003-X}

\bibitem{szabo_2000}
T.~Szabo, J.~Wu, A model for longitudinal and shear wave propagation in
  viscoelastic media.
\newblock Journal of the Acoustical Society of America \textbf{107}, 2437--2446
  (2000)

\bibitem{bryant_1999}
C.~Bryant, D.~McClements, Ultrasonic spectroscopy study of relaxation and
  scattering in whey protein solutions.
\newblock Journal of the Science of Food and Agriculture \textbf{79},
  1754--1760 (1999)

\bibitem{shilov_2002}
V.~Shilov, V.~Sperkach, Y.~Sperkach, A.~Strybulevych, Acoustic relaxation of
  liquid poly(tetramethylene oxide) with hydroxyl and acyl terminal groups.
\newblock Polymer Journal \textbf{34}, 565--574 (2002)

\bibitem{jimenez_2016}
N.~Jim{\'e}nez, F.~Camarena, J.~Redondo, V.~S{\'a}nchez-Morcillo, Y.~Hou,
  E.~Konofagou, Time-domain simulation of ultrasound propagation in a
  tissue-like medium based on the resolution of the nonlinear acoustic
  constitutive relations.
\newblock Acta Acustica \textbf{102}, 876--892 (2016).
\newblock \doi{10.3813/AAA.919002}

\bibitem{holm_2010}
S.~Holm, R.~Sinkus, A unifying fractional wave equation for compressional and
  shear waves.
\newblock Journal of the Acoustical Society of America \textbf{127}, 542--548
  (2010).
\newblock \doi{10.1121/1.3268508}

\bibitem{holm_2011}
S.~Holm, S.~N{\"a}sholm, A causal and fractional all-frequency wave equation
  for lossy media.
\newblock Journal of the Acoustical Society of America \textbf{130}, 2195--2202
  (2011).
\newblock \doi{10.1121/1.3268508}

\bibitem{kinsler}
L.~Kinsler, A.~Frey, A.~Coppens, J.~Sanders, \emph{Fundamentals of Acoustics}
  (Wiley, 2000)

\end{thebibliography}
\end{document}